\documentstyle[aps,prd,eqsecnum]{revtex}
\def\la{\mathrel{\mathchoice {\vcenter{\offinterlineskip\halign
{\hfil$\displaystyle##$\hfil\cr<\cr\sim\cr}}}
{\vcenter{\offinterlineskip\halign{\hfil$\textstyle##$\hfil\cr
<\cr\sim\cr}}}{\vcenter{\offinterlineskip\halign{\hfil$\scriptstyle##$
\hfil\cr<\cr\sim\cr}}}{\vcenter{\offinterlineskip\halign{\hfil$
\scriptscriptstyle##$\hfil\cr<\cr\sim\cr}}}}}
\tighten
\begin{document}
\preprint{IUCAA - 21/95}
\title {Gravitational waves from coalescing binaries: detection strategies
and Monte Carlo estimation of parameters}
\author{R. Balasubramanian \and B.S. Sathyaprakash \and  S. V. Dhurandhar}
\address{Inter-University Centre for Astronomy and Astrophysics, \\
Post Bag 4, Ganeshkhind, Pune 411 007, India}
\date{\today}
\maketitle
\begin{abstract}
The detection of gravitational waves from astrophysical sources is
probably one of the most keenly awaited events in the history of
astrophysics. The paucity of gravitational wave sources and the relative
difficulty in detecting such waves, as compared to those in the
electromagnetic domain, necessitate the development of optimal
data analysis techniques to detect the signal, as well as to extract
the  maximum possible information from the detected signals.
Coalescing binary systems are one of the most promising sources of
gravitational waves. This is due to the fact that such sources are easier
to model and thus one can design detection strategies particularly
tuned to such signals. A lot of attention has been devoted in the
literature studying such techniques and most of the work has
revolved around the Weiner filtering and the maximum likelihood
estimators of the parameters of the binary system. We investigate such
techniques with the aid of differential geometry which provides geometric
insight into  the problem. Such a formalism allows to explore
the  merits and demerits of a detection scheme independent of the
 parameters chosen to represent the waveform. The formalism also
generalises  the problem of choosing an optimal set of templates to detect
a known waveform buried in noisy data. We stress the need for
finding a set of {\em
convenient} parameters for the waveform and show that even after the
inclusion of the second-order post-Newtonian  corrections, the waveform
can essentially be detected by employing a  one-dimensional
lattice of templates.
This would be very useful both for the  purpose of carrying out the
simulations  as well as for the actual detection process.
After setting up such a formalism we carry out a  Monte Carlo simulation
 of the detection process for the initial LIGO/VIRGO configuration for the
first post-Newtonian corrected coalescing binary waveform. We compare
 the results of our
simulations  with the currently available estimates of the accuracies
in the determination of the parameters and the probability
distribution of the maximum likelihood estimators. Our results
suggest that the covariance matrix underestimates, by over a factor of two,
the actual errors in the
estimation of parameters even when the signal-to-noise ratio  is as high
as 20. As only a tiny fraction of the events is expected to be detected
with a signal-to-noise higher than this value, the covariance matrix is
grossly inadequate to describe the errors in the measurement of the
parameters of the waveform. It is found from our Monte Carlo simulations
that the deviations from the covariance matrix are more in
the case of the first post-Newtonian waveform than in the case of the
Newtonian one. Inclusion of higher-order
post-Newtonian corrections introduces new parameters
that are correlated with those at the lower post-Newtonian
waveform. Such correlations are expected to further increase the
discrepancy of the covariance
matrix results with those inferred from Monte Carlo simulations.
Consequently, numerical simulations that take into account post-Newtonian
corrections beyond the first post-Newtonian order are needed in order
to get a clearer picture about the accuracy in the determination of
parameters.
We find that with the aid of the instant of coalescence
the direction to the source can determined more accurately than with the
time of arrival.
\end{abstract}

\mediumtext
\section{Introduction}
Laser interferometric detectors of gravitational waves such as the LIGO
\cite{LIGO} and VIRGO \cite{VIRGO}  are expected to be operational by the
turn of the century.
Gravitational waves from coalescing binary systems of black holes and neutron
stars are relatively `clean' waveforms in the sense that they are easier to
model and for this reason they are amongst the most important
candidate sources for interferometric detectors.
Binary systems are also valuable sources of astrophysical information as
one can probe the universe up to cosmological distances. For instance,
statistical analysis of several binary coalescences enables the estimation
of the Hubble constant to an accuracy better than 10\% \cite {SCH86,Mark}.
Events that produce high signal-to-noise ratio can be potentially used
to observe such non-linear effects, such as gravitational wave
tails, and to put general relativity into test in the strongly non-linear
regime \cite {BS95}.
Due to the weak coupling of gravitational radiation with matter the
signal waveform is in general
very weak and will not stand above the detector noise.
In addition to the on-going efforts to reduce the noise, and hence increase the
sensitivity of the detector, a considerable amount of research activity
has gone into the development of efficient and robust data analysis techniques
to extract signals buried in very noisy data.
For a recent review on gravitational waves from compact objects and their
detection see Thorne \cite{Th95a,Th95b}.

Various data analysis schemes have been suggested for the detection of the
`chirp' waveform from such systems \cite{MS95,Sm87,Th300}. Among  them the
technique of Weiner filtering is the most promising \cite{Th300,HEL,SCH89}.
Briefly, this technique involves correlating the detector output
with a set of templates, each of which is tuned to
detect the signal with a particular set of parameters. This requires the signal
to be known to a high level of accuracy. The fully general relativistic
waveform from a coalescing binary system of stars is as yet unavailable.
In the absence of such an exact solution, there have been
efforts to find solutions perturbatively. Most of the work done in
this area aims at computing the waveform correct to a high degree of
accuracy so that theoretical templates computed based on them will
obtain a large signal-to-noise ratio (SNR) when correlated with the
detector output if the corresponding signal is present.
In general, the number of parameters increases as we incorporate the
higher order corrections.  It is clear that the number of templates depends
upon the number of signal parameters.
As a consequence, the computing power for an on-line analysis will
be greater for a larger number of parameters. In view of this restriction
in computing power it is
necessary to choose the templates in an optimal manner.
This paper in part deals with this question. Investigations
till now have been restricted to either choosing a finite subset of the signal
space as templates \cite{SD91,DS94} or choosing templates from the
`Newtonian' or the `first post-Newtonian' family of waveforms
\cite{BD94,KKT94,A95,KKS95}.  We generalize this problem by using the language
of differential geometry. We show that it is unnecessary to restrict
oneself to templates that are matched exactly to any particular signal.

Differential geometry has been used in statistics before (see \cite{Am}
and references therein) and the
standard  approach is to treat a set of parameterised probability distributions
corresponding to a particular statistical model as a manifold.  The
parameters of the distribution serve as coordinates on this manifold.
In statistical theory one frequently comes across the Fisher information matrix
whose inverse gives a lower bound for the errors in the estimation of the
parameters of a distribution.
The Fisher information matrix turns out to be a
very natural metric on the  manifold of probability distributions
and this metric can be used profitably
in understanding the properties of a particular statistical model. Here in our
paper we treat the set of coalescing binary signals corresponding to
various parameters of the binary as a submanifold in the linear space of all
detector outputs. We show in this paper that both the above mentioned
manifolds are equivalent as far as their metrical properties are concerned.
The geometric approach turns out to be useful not only in
clarifying various aspects of signal analysis but also helps us to pose the
question of optimal detection in a more general setting.

Once a signal has been detected we can estimate the parameters of the binary.
We assume that the parameters of the signal are the same as those of
the template with which the maximum correlation is obtained. The errors
involved in such an estimation has been worked out by several
authors \cite{BS95,KKS95,F92,FC93,CF94,BS94,Kr,KLM93,PW95,JKKT},
for the case of `high' SNR and for the Newtonian and post-Newtonian
waveforms using a single and a network of detectors.
For the case of low SNRs one has to resort to numerical simulations.
We have started a project to carry out exhaustive
numerical simulations specifically designed to compute the errors in the
estimation of parameters and covariances among them at various post-Newtonian
orders, for circular and eccentric orbits, with and without spin effects and
for different optical configurations of the interferometer.
In this paper we report the results for the case of the initial
LIGO configuration taking
into account only the first post-Newtonian corrections and
assuming circular orbits. Going beyond this needs tremendous amount of
computing power which is just becoming available.

The rest of the paper is organized as follows.
In section \ref{waveforms} we describe the waveform from a coalescing binary
system at various post-Newtonian orders. We introduce, following  \cite{Sat94},
a set of parameters called `chirp times'. These parameters are found to
be very convenient when we carry out Monte Carlo simulations.
It turns out that the covariance matrix  is independent  of these parameters
and hence it is sufficient to carry out the simulations only  for a particular
set of parameters. In section \ref{sm} we develop a geometric interpretation
of signal analysis. We begin by introducing a metric on the manifold
from a scalar product, which comes naturally from the theory of matched
filtering and
then show that this metric is the same as the one used by Amari \cite{Am}.
Using the geometric  approach we address the question of optimal filter
placement and show that for the purpose of detection it is optimal to
choose the templates outside the signal manifold.
The covariance matrix of errors and covariances is shown to be the inverse
of the metric on the manifold. In section \ref{sec_espar} we  discuss
the results of our simulations and
compare the numerically obtained values and those suggested by
the covariance matrix. We find substantial discrepancies in the predictions
of the two methods. It is believed that the coalescing binary waveform
shuts off abruptly at the onset of the plunge orbit. This has a major effect on
the computations of the covariance matrix as well as on the Monte Carlo
simulations.  We discuss the effects of higher post-Newtonian corrections
to the waveform. We also emphasisize the use of the instant of coalescence
as a parameter in order to determine the direction to the
source rather than the time of arrival \cite {BSD95}.
Finally in section \ref{sec_con} we summarise our results and indicate future
directions.

\section{Coalescing Binary Waveforms}
\label{waveforms}

For the purpose of constructing templates for on-line detection,
it is sufficient to work with the so called {\it restricted} post-Newtonian
gravitational waveform. In this approximation
the post-Newtonian corrections are incorporated only in the phase of the
waveform, while ignoring corresponding corrections to the amplitude
\cite {3mn}. Consequently, the restricted post-Newtonian waveforms
only contain the dominant frequency equal to twice the orbital frequency
of the binary computed up to the relevant order.
In the restricted post-Newtonian approximation the gravitational
waves from a binary system of stars, modeled as point masses orbiting
about each other in a circular orbit, induce a strain $h(t)$ at the
detector given by
\begin {equation}
h(t) = A (\pi f(t) )^{2/3} \cos \left [\varphi (t) \right ],
\label {wave}
\end {equation}
where $f(t)$ is the instantaneous gravitational wave frequency,
the constant $A$ involves the distance to the binary, its reduced
and total mass, and the antenna pattern of the detector \cite{Th300},
and the phase of the waveform $\varphi (t)$ contains several pieces
corresponding to different post-Newtonian contributions which
can be schematically written as
\begin {equation}
\varphi(t) = \varphi_0(t) + \varphi_1(t) + \varphi_{1.5}(t) + \ldots.
\label {phase}
\end {equation}
Here $\varphi_0(t)$ is the dominant Newtonian part of the phase
and $\varphi_n$ represents the $n$th order post-Newtonian
correction to it. In the quadrupole approximation we include
only the Newtonian part of the phase given by \cite {Th300}
\begin {equation}
\varphi (t) = \varphi_0 (t) =
{16 \pi f_a \tau_0 \over 5}
\left [ 1 - \left ({f\over f_a}\right )^{-5/3} \right] + \Phi,
\label {phaseN}
\end {equation}
where $f(t)$ is the instantaneous Newtonian gravitational
wave frequency given implicitly by
\begin {equation}
t - t_a = \tau_0
\left [ 1 - \left ( {f \over f_a} \right )^{-8/3} \right ],
\label {frequencyN}
\end {equation}
$\tau_0$ is a constant having dimensions of time given by
\begin {equation}
\tau_0 = {5 \over 256} {\cal M}^{-5/3} (\pi f_a)^{-8/3},
\label {NCT}
\end {equation}
and $f_a$ and $\Phi$ are the instantaneous gravitational
wave frequency and the phase of the signal, respectively, at $t=t_a.$
The time elapsed starting from an epoch when the
gravitational wave frequency is $f_a$ till the epoch when
it becomes infinite will be referred to as
the {\it chirp time} of the signal.
In the quadrupole approximation $\tau_0$ is the chirp time.
The Newtonian part of the phase
is characterised by three parameters:
(i) the {\it time of arrival} $t_a$
when the signal first becomes {\it visible} in the detector,
(ii) the {\it phase} $\Phi$ of the signal at the time of arrival
and (iii) the {\it chirp mass} ${\cal M} = (\mu^3 M^2)^{1/5},$
where $\mu$ and $M$ are the reduced and the total mass
of the binary, respectively.
At this level of approximation two coalescing binary signals of
the same chirp mass but
of different sets of individual masses would be degenerate and thus exhibit
exactly the same time evolution.  This
degeneracy is removed when post-Newtonian corrections are included.

When post-Newtonian corrections are included
the parameter space of waveforms acquires an extra dimension.
In this paper we show that even when
post-Newtonian corrections up to relative order $c^{-4},$
where $c$ is the velocity of light, are included in the phase of
the waveform it is possible to make a judicious choice of the
parameters so that the parameter space essentially remains only
three dimensional as far as the detection problem is concerned.
It should, however, be noted that
the evolution of the waveform must be known to a reasonably high
degree of accuracy and that further off-line analysis would be
necessary to extract useful astrophysical information.

With the inclusion of corrections up to second post-Newtonian order the
phase of the waveform becomes \cite {BDI}
\begin {equation}
\varphi (t) = \varphi_0 (t) + \varphi_1 (t)  + \varphi_{1.5} (t)
+ \varphi_2 (t)
\label {phasetotal}
\end {equation}
where $\varphi_0(t)$ is given by (\ref {phaseN}) and the various
post-Newtonian contributions are given by
\begin {equation}
\varphi_1 (t) =
4 \pi f_a \tau_1 \left [ 1 - \left ( {f\over f_a} \right )^{-1} \right ]
\label {phase1PN}
\end {equation}
\begin {equation}
\varphi_{1.5} (t) =   - 5 \pi f_a \tau_{1.5}
\left [ 1 - \left ( {f\over f_a} \right )^{-2/3} \right ]
\label {phase1.5PN}
\end {equation}
and
\begin {equation}
\varphi_2 (t) =   8 \pi f_a \tau_2
\left [ 1 - \left ( {f\over f_a} \right )^{-1/3} \right ].
\label {phase2PN}
\end {equation}
Now $f(t)$ is the instantaneous gravitational wave frequency correct up to
second post-Newtonian order given implicitly by
\begin {equation}
t - t_a =
\tau_0 \left [ 1 - \left ( {f \over f_a} \right )^{-8/3} \right ] +
\tau_1 \left [ 1 - \left ( {f \over f_a} \right )^{-2} \right ] -\\
 \tau_{1.5} \left [ 1 - \left ( {f \over f_a} \right )^{-5/3} \right ]
+ \tau_2 \left [ 1 - \left ( {f \over f_a} \right )^{-4/3} \right ].
\label {frequencyPN}
\end {equation}
In the above equations $\tau$'s are constants having dimensions of time
which depend only on the two masses of the stars and the lower frequency
cutoff of the detector $f_a$.
The total chirp time now consists of four pieces: the Newtonian contribution
$\tau_0$ is given by (\ref {NCT}) and the various post-Newtonian contributions
are
\begin {equation}
\tau_1 = {5 \over 192\mu (\pi f_a)^2} \left ({743\over 336} + {11\over 4} \eta
\right ),
\end {equation}
\begin {equation}
\tau_{1.5} = {1\over 8 \mu} \left ( m \over \pi^2 f_a^5 \right)^{1/3}
\end {equation}
and
\begin {equation}
\tau_2 = {5\over 128 \mu} \left ({m\over \pi^2 f_a^2}\right )^{2/3}
\left ({3058673 \over 1016064} + {5429\over 1008} \eta +
{617\over 144}\eta^2\right)
\end {equation}
where $\eta=\mu/M.$
The phase (\ref {phasetotal})
contains the reduced mass $\mu$ in addition to the chirp mass $\cal M.$
Taking ($\cal M$, $\eta$) to be the post-Newtonian mass parameters
the total mass and the reduced mass are given by
$M = {\cal M} \eta^{-3/5},$ $\mu = {\cal M} \eta^{2/5}.$
Note that in the total chirp time $\tau$ of the signal
the 1.5 post-Newtonian
contribution appears with a negative sign thus shortening the epoch of
coalescence.

With the inclusion of higher order post-Newtonian corrections a
chirp template is characterised by a set of
four parameters which we shall collectively denote by $\lambda^\mu,$
$\mu=1,\ldots, 4.$
At the first post-Newtonian approximation instead of working with the
parameters $\lambda^\mu =\{t_a,~\Phi,~{\cal M},~\eta\}$ we can equivalently
employ the set $\{t_a,~\Phi,~\tau_0,~\tau_1\}$ for the purpose of
constructing templates.
This, as we shall see later, has some advantageous. However,
at post-Newtonian orders beyond the first we do not have a unique
set of chirp times to work with.

The parameters $t_a$ and
$\Phi$ are {\it kinematical} that fix the origin of the
measurement of time and phase, respectively, while the Newtonian and the
post-Newtonian chirp times
are {\it dynamical} parameters in the sense that they dictate the
evolution of the phase and the amplitude of the signal.
It may be mentioned at this stage that in most of the literature
on this subject authors use the set of parameters $\{t_C,~\Phi_C,~{\cal M},~
\eta\}$ where $t_C$ is the instant of coalescence and $\Phi_C$ is
the phase of the signal at the instant of coalescence. In terms of
the chirp times we have introduced, $t_C$
is the sum of the total chirp time and the time of arrival
and $\Phi_C$ is a combination of the various chirp times and $\Phi:$
\begin {equation}
t_C = t_a + \tau_0 + \tau_1 - \tau_{1.5} + \tau_2; \quad
\end {equation}
\begin {equation}
\Phi_C = \Phi + {16\pi f_a \over 5} \tau_0 +
4 \pi f_a \tau_1 - 5\pi f_a \tau_{1.5} + 8\pi f_a \tau_2.
\end {equation}

In the stationary phase approximation
the Fourier transform of the restricted first-post-Newtonian
chirp waveform for positive frequencies is given by \cite
{Th300,SD91,FC93,CF94}
\begin {equation}
\tilde h (f) = {\cal N} f^{-7/6} \exp \left
[i\sum_{\mu=1}^6\psi_\mu(f)\lambda^\mu
- i {\pi \over 4} \right ]
\label {FT}
\end {equation}
where
$${\cal N} = A\pi^{2/3} \left ( {2\tau_0 \over 3}\right )^{1/2} f_a^{4/3} $$
is a normalisation constant, $\lambda^\mu,$ $\mu=1,\ldots, 6,$ represent
the various post-Newtonian parameters
\begin {equation}
\lambda^\mu = \left \{t_a, \Phi, \tau_0, \tau_1, \tau_{1.5}, \tau_2
\right \},
\end {equation}
and
\begin {eqnarray}
\label {eqs1}
\psi_1 & = & 2\pi f, \\
\psi_2 & = & -1, \\
\psi_3 & = & 2 \pi f  -{ 16 \pi f_a \over 5}+ {6\pi f_a \over 5}
\left ( {f\over f_a} \right )^{-5/3},\\
\psi_4 & = & 2\pi f - 4\pi f_a + 2\pi f_a \left ({f\over f_a}\right)^{-1},\\
\psi_5 & = &-2\pi f + 5\pi f_a - 3\pi f_a \left ({f\over f_a}\right)^{-2/3},\\
\psi_6 & = & 2\pi f - 8\pi f_a + 6\pi f_a \left ({f\over f_a}\right)^{-1/3}.
\label {eqs2}
\end {eqnarray}
For $f<0$ the Fourier transform is computed using the identity $\tilde
h(-f) = \tilde h^*(f)$ obeyed by real functions $h(t).$ In addition to the
above mentioned
parameters we shall introduce an amplitude parameter $\cal A$ in section
\ref{sm}.

\section{A geometric approach to signal analysis}
\label{sm}
In this section we apply the techniques of differential geometry to the
problem of detecting weak signals embedded in noise.  In section \ref{sms} we
introduce the concept of the signal manifold and elaborate on the relationship
of our approach with that of Amari \cite{Am}. Our discussion of the
vector space of all detector outputs is modeled after the discussion
given in \cite{SS94}.
In section \ref{detect} we deal with the problem of choosing a set of
filters for on-line analysis which would optimise the task of detection of
the signal. In section \ref{2pnsec} we deal with the dimensionality of the
chirp manifold when we incorporate higher order post-Newtonian corrections.
It is found, that due to covariances between the parameters, it is possible
to introduce an effective dimension which is less than the dimension of the
manifold.
This has very important implications for the detection problem.

\subsection{Signal manifold}
\label{sms}
The output of a gravitational wave detector such as the
LIGO, will comprise of data segments, each of
duration $T$ seconds, uniformly sampled with a sampling interval of
$\Delta$, giving the number of samples in a single data train to be
$N = T/\Delta$. Each data train can be considered as a $N$-tuple
$(x^0,x^1,\ldots,x^{N-1})$\,  $x^i$ being the  value of the output of the
detector at time $i\Delta$. The set of all such $N$ tuples constitutes a
$N$-dimensional  vector space $\cal V$  where the addition of two vectors is
accomplished by the addition of corresponding time samples.
For later convenience we allow each sample to take complex values. A natural
basis for this vector space is  the {\em time basis} ${\bf e}_m^i =
\delta^i_m$ where $m$ and $i$ are the vector and component indices
respectively. Another basis which we shall use extensively is the Fourier
basis which is related to the time basis by a unitary transformation
$\hat U$:
\begin{eqnarray}
 {\bf\tilde e}_m &=& \hat U^{mn}{\bf e}_n = \frac{1}{\sqrt N}
\sum\limits_{n=0}^{N-1}
{\bf e}_n\exp
\left[\frac{2\pi i m n}{N}\right],\\
 {\bf e}_m &=& \hat U^{\dagger mn}{\bf\tilde e}_n = \frac{1}{\sqrt N}
 \sum\limits_{n=0}^{N-1}{\bf\tilde e}_n\exp
\left[-\frac{2\pi i m n}{N}\right].
\end{eqnarray}
All vectors in $\cal V$ are shown in boldface, and the Fourier basis vectors
and
components of vectors in the Fourier basis are highlighted with a `tilde'.

In the continuum case each data train can be expanded in a Fourier series
and will contain a finite number of terms in the expansion, as the
output will be band limited. The expansion is
carried out over the exponential functions $\exp\left(2\pi i m t/T\right)$
which are precisely the Fourier basis vectors defined above. Though the index
$m$ takes both positive and negative values corresponding to positive and
negative frequencies it is both, possible and convenient to
allow $m$ to take only positive values \cite{NUM}. Thus the
vector space $\cal V$ can be considered as being spanned by the $N$ Fourier
 basis vectors implying immediately that the number of independent vectors in
the time basis to be also $N$. This is the content of the Nyquist theorem
which states that it is sufficient to sample the data  at a
frequency which is  twice as large as the bandwidth of a real valued signal,
where the bandwidth
refers to the range of positive frequencies over which the signal spectrum is
non zero.
This factor of two does not appear in the vector space picture as we allow
in general for complex values for the components in the time basis.

A gravitational wave  signal from a coalescing binary system
can  be characterised by a set of parameters
$\bbox{\lambda} = (\lambda^0,\lambda^1,\ldots,\lambda^{p-1})$ belonging to some
open set of the $p$-dimensional real space $R^p$.
The set of such signals
${\bf s}(t;\bbox{\lambda})$ constitutes a $p$-dimensional manifold $\cal S$
which is
embedded in the vector space $\cal V$. The parameters of the binary act as
coordinates on the manifold. The basic problem of signal analysis is thus to
 determine whether the detector output vector $\bf x$ is
the sum of a signal vector and a noise vector, ${\bf x} = {\bf s} + {\bf n}$,
or just the noise vector, ${\bf x} = {\bf n}$, and furthermore to identify
which
 particular signal vector, among all possible.
One would also like to estimate the errors in such a measurement.

In the absence of the signal the output will contain
only noise drawn from a stochastic process which can be described by a
probability distribution on the vector space $\cal V$.
The covariance matrix of the noise $C^{ij}$ is defined as,
\begin{equation}
C^{ij} = \overline{n^in^{*j}},
\end{equation}
where an * denotes complex conjugation and an overbar denotes an
average over an ensemble. If the noise is assumed to be stationary and ergodic
then
there exists a noise correlation function $K(t)$ such that $C_{ij} =
K(|i-j|\Delta)$. In the Fourier basis it can be shown that the components
of the noise vector are statistically independent \cite{HEL} and  the
covariance matrix in the Fourier
 basis will contain only diagonal terms whose values will be strictly positive:
 $\tilde C_{ii} = \overline{\tilde n^i\tilde n^{*i}}$.
 This implies that the covariance matrix
has strictly positive eigenvalues. The diagonal elements of this
matrix $\tilde C_{ii}$ constitute the discrete representation of the power
spectrum of the noise $S_n(f)$.

We now discuss how the concept of matched filtering can be used to induce a
metric
on the signal manifold. The technique of matched filtering involves correlating
the
detector output with a bank of filters each of which is tuned to detect the
gravitational wave from a binary system with a particular set of parameters.
The output of the filter, with an impulse response $\bf q$, is given in the
discrete case
as
\begin{equation}
\label{corr}
c_{(m)}  = \frac{1}{\sqrt N}\sum\limits_{n=0}^{N-1}\tilde x^n \tilde q^{*n}
\exp\left[-2\pi imn/N\right].
\end{equation}
The SNR ($\rho$) at the output is defined to be the mean of $c_{(m)}$ divided
by the square root of its variance:
\begin{equation}
\label{rho1}
\rho \equiv \frac{\overline{c_{(m)}}}{\left[\overline{\left(c_{(m)} -
\overline{c_{(m)}}\right)^2}
\right]^{1/2}}.
\end{equation}
By maximising $\rho$ we can obtain the expression
for the matched filter $\bf q_{(m)}$ matched to a particular signal ${\bf
s}(t;\lambda^\mu)$ as
\begin{equation}
\label{matched}
\tilde q^n_{(m)}(\lambda^\mu) = \frac{\tilde s^n(\lambda^\mu)\exp\left[2\pi
imn/N\right]}{\tilde C_{nn}},
\end{equation}
where $\rho$ has been maximised at the $m^{th}$ data point at the output and
where $\mu = 1,2,\ldots,p$
where $p$ is the number of parameters of the signal. We now introduce a scalar
product in $\cal V$.
For any two vectors $\bf x$ and $\bf y$,
\begin{equation}
\label{scal}
\left\langle{\bf x},{\bf s}\right\rangle =
\sum\limits_{i,j=0}^{N-1}C^{-1}_{ij}x^iy^j =
\sum\limits_{n=0}^{N-1}\frac{\tilde x^n \tilde s^{*n}(\lambda^\mu)}{\tilde
C_{nn}}.
\end{equation}
In terms of this scalar product, the output of the matched filter $\bf q$,
matched to a signal
${\bf s}(\lambda^\mu)$, can be written as,
 \begin{equation}
\label{eqsc}
c_{(m)}(\lambda^\mu) = \frac{1}{\sqrt N}\left\langle{\bf x},{\bf
s}\right\rangle.
\end{equation}
As $\tilde C_{ii}$ is strictly positive the scalar product defined
is positive definite.
The scalar product defined above on the vector space $\cal V$ can be used to
define a norm
on $\cal V$ which in turn can be used to induce a metric on the manifold. The
norm
of a vector $\bf x$ is defined as $\|{\bf x}\| = \left\langle{\bf x},{\bf x}
\right\rangle^{1/2}$.
The  ratio for the matched filter can  be
calculated to give $\rho = \left\langle{\bf s},{\bf s}\right\rangle^{1/2}$.
The norm of the noise vector will be a random variable $\left\langle{\bf
n},{\bf n}\right\rangle^{1/2}$
with a mean of $\sqrt{N}$ as can be seen by writing the expression for the norm
of the noise
vector and subsequently taking an ensemble average.

The distance between two points infinitesimally separated on $\cal S$ can be
expressed as
a quadratic form in
the differences in the values of the parameters at the two points:
\begin{eqnarray}
g_{\mu\nu}d\lambda^\mu d\lambda^\nu &\equiv& \|{\bf s}(\lambda^\mu +
d\lambda^\mu) -
{\bf s}(\lambda^\mu)\| =  \|\frac{\partial{\bf
s}}{\partial\lambda^\mu}d\lambda^\mu\|\\
&=& \left\langle\frac{\partial{\bf s}}{\partial\lambda^\mu},
\frac{\partial{\bf s}}{\partial\lambda^\nu}\right\rangle d\lambda^\mu
d\lambda^\nu
\end{eqnarray}
The components of the metric in the coordinate basis are seen to be the scalar
products
of the coordinate basis vectors of the manifold.

Since the number of correlations we can perform on-line is finite, we cannot
have a
 filter corresponding
to every signal. A single filter though matched to a particular signal will
also
detect signals in a small neighbourhood of that signal but with a loss
in the SNR. The metric on the manifold quantifies the drop in the
 correlation
in a neighbourhood of the signal chosen.
 Taking the output vector to be $\bf x$ and two signal vectors
$\bf s(\bbox{\lambda})$ and $\bf s(\bbox{\lambda}+d\bbox{\lambda})$
 and using Schwarz's inequality we have,
\begin{eqnarray}
\left\langle{\bf x},{\bf s(\bbox{\lambda}+d\bbox{\lambda})}\right\rangle -
\left\langle{\bf x},{\bf s(\bbox{\lambda})}\right\rangle
= \left\langle{\bf x},{\bf s}(\bbox{\lambda}+d\bbox{\lambda}) - {\bf
s}(\bbox{\lambda})
\right\rangle
&\le& \|{\bf x}\|\|{\bf s}(\bbox{\lambda} + d\bbox{\lambda}) - {\bf
s}(\bbox{\lambda})\|\\
&=& \|{\bf x}\|g_{\mu\nu}d\lambda^\mu d\lambda^\nu.
\end{eqnarray}
As is apparent the drop in the correlation can be related to the metric
distance on the manifold between the two signal vectors.

We now discuss Amari's \cite{Am} work in the context of using differential
geometry in statistics and  elaborate on the relationship with the approach
we have taken. The set of parametrised probability distributions corresponding
to a statistical model constitute a manifold. The parameterized probability
distributions in the context of signal analysis of gravitational waves from
coalescing binaries are the ones which specify the probability that the output
vector will lie in a certain region of the vector space $\cal V$ given that a
signal
${\bf s}(t;\bbox{\lambda})$ exists in the output which we denote as $p({\bf x}|
{\bf s}(t;\bbox{\lambda}))$
 Since it is not our intention
to develop Amari's approach any further we will be brief and will make
all the mathematical assumptions such as infinite differentiability of
functions,
interchangeability of the differentiation and expectation value operators, etc.

	The set of probability distributions $p({\bf x}|{\bf s}(\bbox{\lambda}))$
where $\bbox{\lambda} \in R^p$, constitutes a manifold $\cal P$ of dimension
$p$.
  At every point
on this manifold we can construct a tangent space $T^0$ on which we can define
the
coordinate basis vectors as ${\bf\partial}_\mu =
\frac{\bf\partial}{{\bf\partial}\lambda^\mu}$.
Any vector $\bf A$ in this tangent space can be written as a linear combination
of these coordinate basis vectors.
We now define $p$ random variables $\sigma_\mu =
-\frac{\partial}{\partial\lambda^\mu}\log(p({\bf x}|{\bf s}(\bbox{\lambda})))$.
It can easily be shown that $\overline{\sigma_\mu} = 0$. We assume that these
$p$ random
variables are linearly independent. By taking all possible linear combinations
of these random
variables we can construct another linear space $T^1$. Each vector
$\bf B$ in $T^1$ can be written as ${\bf B} = B^\mu\sigma_\mu$. The two vector
spaces
$T^0$ and $T^1$
are isomorphic to each other, which can be shown explicitly  by
making the correspondence  $\sigma_\mu \leftrightarrow {\bf\partial}_\mu$. The
vector space
$T^1$ has a natural inner product defined on it which is the covariance
matrix of the $p$ random variables $\sigma_\mu$. This scalar product can be
carried
over to $T^0$ using the correspondence stated above. The metric on the
manifold can be defined by taking the scalar product of the coordinate basis
vectors
\begin{equation}
\label{met}
g_{\mu\nu} = \left\langle \partial_\mu,\partial_\nu\right\rangle =
\overline{\sigma_\mu\sigma_\nu}
\end{equation}
In statistical theory the above matrix $g_{\mu\nu}$ is called the Fisher
information
matrix. We will also denote the Fisher matrix, as is conventional, as
$\Gamma_{\mu\nu}$.
It is clearly seen that orthogonality between vectors in the tangent
space of the manifold is related to statistical independence of random
variables
in $T^1$.

If we take the case of Gaussian noise the metric defined above
is identical to the one obtained on the signal manifold by matched filtering.
Gaussian noise can be described by the distribution,
\begin{equation}
\label{ndis}
p({\bf n}) = \frac{\exp\left[
	-\frac{1}{2}\sum\limits_{j,k=0}^{N-1}{C^{-1}_{jk} n^j n^{k*}}\right]}
{ \left[\ (2\pi)^N \det\left[C_{ij}\right]\ \right]^{1/2}} = \frac{\exp\left[
	-\frac{1}{2}\sum\limits_{j,k=0}^{N-1}{\tilde C^{-1}_{jk} \tilde n^j
\tilde n^{k*}}\right]} { \left[\ (2\pi)^N \det\left[\tilde C_{ij}\right]\
\right]^{1/2}}
= \frac{\exp\left[
	-\frac{1}{2}\sum\limits_{i=0}^{N-1}{\frac{\tilde n^i
\tilde n^{i*}}{\tilde C_{ii}}}\right]} { \left[\ (2\pi)^N \det\left[\tilde
C_{ij}\right]\ \right]^{1/2}},
\end{equation}
where in the last step we have used the diagonal property of the matrix $\tilde
C^{ij}$
which implies that $\tilde C^{-1}_{ii} = 1/\tilde C_{ii}$.

As the noise is additive $p({\bf x}|{\bf s}(\bbox{\lambda}))$ can be written
as  $p({\bf x} - {\bf s}(\bbox{\lambda}))$.  Assuming Gaussian noise we can
write the
expressions for the random variables $\sigma_i$ as,
\begin{equation}
\label{sig}
\sigma_\mu =
\left(\frac{1}{2}\right)\frac{\partial}{\partial\lambda^\mu}\left\langle{\bf
x}-{\bf s}(\bbox{\lambda}),
{\bf x} - {\bf s(\bbox{\lambda})}\right\rangle = \left\langle{\bf
n},\frac{\partial}{\partial\lambda^\mu}
{\bf s}(\bbox{\lambda})\right\rangle,
\end{equation}
where in the last step we have used ${\bf x} = {\bf s}(\bbox{\lambda})+{\bf
n}$. The covariance matrix
for the random variables $\sigma_\mu$ can be calculated to give
\begin{equation}
\label{sigco}
\overline{\sigma_\mu\sigma_\nu} =
\left\langle\frac{\partial}{\partial\lambda^\mu}{\bf s}(\bbox{\lambda}),
\frac{\partial}{\partial\lambda^\nu}{\bf s}(\bbox{\lambda})\right\rangle,
\end{equation}
which is the same metric as defined over the signal manifold. Thus, both the
manifolds
$\cal S$ and $\cal P$ are identical with respect to their metrical properties.
We will henceforth restrict  our attention to the signal manifold  $\cal S$.

For the purpose of our analysis we will choose a minimal set of parameters
characterizing the gravitational wave signal from a coalescing binary. We
consider
only the first post-Newtonian corrections. In section \ref{waveforms} we have
already
introduced the four parameters $\lambda^\mu = \{t_a,\Phi,\tau_0,\tau_1\}$. We
now introduce
an additional parameter for the amplitude and call it $\lambda^0 = \cal A$.
The signal can now be written as $\tilde s(f;\bbox{\lambda}) =
{\cal A} h(f;t_a,\Phi,\tau_0,\tau_1)$, where,
$\bbox{\lambda} \equiv \{{\cal A},t_a,\Phi,\tau_0,\tau_1\}$. Numerically
the value of the parameter $\cal A$ will be the same as that of the SNR
obtained for the matched filter. We can decompose the signal manifold into a
manifold containing normalised chirp waveforms and a one-dimensional manifold
corresponding
to the parameter $\cal A$.  The normalised chirp manifold can therefore be
parameterized by
$\{t_a,\Phi,\tau_0,\tau_1\}$.
This parameterization is useful as the coordinate basis vector
$\frac{\partial}{\partial
{\cal A}}$ will  be orthogonal to all the other basis vectors as will be seen
below.

In order to compute the metric, and equivalently the Fisher information matrix,
we use the continuum version of the scalar product as given in \cite{F92},
except that we
use the two sided power spectral density.
This has the advantage of showing clearly
the range of integration in the frequency space though we get the same result
using the discrete
 version of the scalar product. Using the definition of the scalar product we
get
\begin {equation}
g_{\mu\nu} = \int_{f_a}^\infty {df\over S_n(f)}
{\partial \tilde s(f; \bbox{\lambda})\over \partial \lambda^\mu}
{\partial \tilde s^* (f; \bbox{\lambda}) \over \partial \lambda^\nu}
+ {\rm c.c.}
\label{gamma2}
\end {equation}
Recall that in the
stationary phase approximation the Fourier transform of
the coalescing binary waveform is given by $\tilde h(f) = {\cal N} f^{-7/6}
\exp \left [ i\sum_\mu\psi_\mu(f) \lambda^\mu \right]$ and $\tilde s(f) = {\cal
A}\tilde h(f)$,
 where $\psi_\mu(f)$ are given
by equations (\ref {eqs1}-\ref{eqs2}), $\mu=1,\ldots, 4$ and
$\lambda^\mu = \{t_a,~\Phi,~\tau_0,~\tau_1\}.$
Note, in particular, that in the phase of the waveform
the parameters occur linearly thus enabling a very concise expression
for the components of  $g_{\mu\nu}.$ The various partial
derivatives are given by
\begin {equation}
{\partial \tilde s (f;\bbox{\lambda}) \over \partial {\lambda^\mu}} =
i \psi_\mu (f) \tilde s(f;\bbox{\lambda}),
\label{derivatives}
\end {equation}
where we have introduced $\psi_0= -i/{\cal A}.$
On substituting the above expressions for the partial derivatives in
eq. (\ref {gamma2}) we get,
\begin {equation}
g_{\mu\nu} =  \left < \psi_\mu h, \psi_\nu h \right >
=  2 \int_{f_a}^\infty {{\psi_\mu(f) \psi^*_\nu(f) \left |\tilde h(f)\right|^2}
\over S_n(f)} df
\label{gamma3}
\end {equation}
The above definition of the amplitude parameter ${\cal A},$
as in Culter and Flannagan \cite {CF94}, disjoins the amplitude of the waveform
from the rest of the parameters. Since $\psi_0$ is pure imaginary and
$\psi_\mu$'s are real, it is straightforward to see from eq. (\ref {gamma3})
that
\begin {equation}
g_{00} = 1 \mbox{  and  } g_{0\mu} = 0; \ \ \mu=1,\ldots,4.
\end {equation}
The rest of the components $g_{\mu\nu}$ are seen to be independent of all the
parameters
except $\cal A$ {\em i.e.} $g_{\mu\nu} \propto {\cal A}^2$. As ${\cal A}$ is
unity for the
normalised manifold the metric on the normalised manifold is flat. This implies
not only that the manifold is intrinsically flat (in the stationary phase
approximation)
but also that the coordinate system used is Cartesian. If instead of the chirp
time
$\tau_0$ we use the parameter $\cal M$ then the metric coefficients will
involve that parameter
and the coordinate system will no longer remain Cartesian.

\subsection{Choice of filters }
\label{detect}

We now use the formalism developed to tackle the issue of optimal filter
placement. Till
now, it has been thought necessary to use a finite subset of the set of chirp
signals
as templates for detection. We show that this is unduly restrictive. We suggest
a procedure by which the
 detection process can be made more `efficient' by moving the filters out of
the manifold.
It must be emphasised that the algorithm presented below is both, simplistic
and quite adhoc and
is not necessarily the best.
Moreover, we have implemented the algorithm only for the Newtonian case where
the computational
requirements are not very heavy. The signal manifold corresponding to higher
post-Newtonian
corrections will be a higher dimensional manifold and here the computational
requirements
will be substantial. The choice of optimal filters which span the manifold will
then be crucial.

Detection of the coalescing binary signal  involves computing the scalar
product
of the output of the detector with the signal vectors. Subsequently one would
have to maximise
the correlations over the parameters and the number so obtained would serve as
the
statistic on the basis of which
 we can decide whether a signal is present in the given data train.
Geometrically, this maximization
corresponds to minimizing the angle between the output vector and the vectors
corresponding
to the normalised signal manifold. Using the cosine formula,
\begin{equation}
\cos(\theta) =
\frac{\left<{\bf s}(\bbox{\lambda}),{\bf x}\right>}{\|{\bf x}\|\|{\bf
s}(\bbox{\lambda})\|}
= \frac{\|{\bf s}(\bbox{\lambda})\|^2 + \|{\bf x}\|^2 -
\|{\bf x} - {\bf s}(\bbox{\lambda})\|^2}
{2\|{\bf x}\|\|{\bf s}(\bbox{\lambda})\|},
\end{equation}
it is clear that as $\|{\bf s}(\bbox{\lambda})\|$ is unity for the vectors
belonging to the
normalised signal manifold,
 maximising the scalar product is equivalent to minimising
$ \|{\bf x} - {\bf s}(\bbox{\lambda})\|$ which is the
distance between the tip of the output vector and the manifold.

Given the constraints of computational
power one would be able to evaluate only a finite number of these
scalar products, say $n_F$, in a certain amount of time depending on the data
train length
 and the padding factor.
It is therefore necessary to be able to choose the $n_F$ filters in such a
manner that
the detection probability is maximal.
We will need efficient on-line data analysis for two reasons:
(i) To isolate those
data trains which have a high probability of containing a signal and (ii) to
determine the parameters
of the binary early on during the inspiral and to use them for dynamical
recycling techniques.

Due to the finiteness
of the filter spacing the signal parameters will in general not correspond
to any of the $n_F$ filters chosen and this will lead to a drop in the maximum
possible
correlation.
Till now attention has been focussed on identifying a set of optimal set of
filters which are
a discrete subset of the manifold.
 If detection is the sole purpose then the differential geometric
picture suggests that confining the filter vectors
to the signal manifold is an unnecessary restriction and in fact non optimal.
Thus it
is worthwhile to explore making a choice of filters outside the manifold.
The filter vectors will thus belong to $\cal V$ but will not, in general,
correspond to any signal.
It is, of course, true that we are sacrificing on the maximum possible
correlation obtainable (when
the signal's parameters coincide with those of the filter). Thus the problem
essentially is
to select $n_F$ filter vectors which optimize the detection the efficiency of
which depends
upon the properties of the manifold.

In general a single filter vector would have to pick up signals over a region
of the manifold. The extent
of this region is determined by fixing a threshold on the correlation between
the filter and
any signal in the region. We will denote this threshold by $\eta$, where $\eta$
takes a value
which is close to, but less than unity. The typical value suggested for $\eta$
is $\sim 0.8$,
\cite{SD91}.
For a given filter $\bf q$ and a threshold
$\eta$  the region on the manifold corresponding to the filter will be denoted
as
${\cal S}_{\bf q}(\eta)$, where ${\cal S}_{\bf q}(\eta) \subset \cal S$.
 Geometrically this region is the intersection of an open ball in $\cal V$ of
radius
$2^{1/2}(1-\eta)^{1/2}$ (using the distance defined by the scalar product),
with center $\bf q$, and the manifold.
The $n_F$ filters taken together would have to `span' the manifold which means
that the union of
the regions covered by each filter would be the manifold itself {\em i.e.},
$\bigcup_{\bf q} {\cal S}_{\bf q}(\eta) = \cal S$.
If the filter $\bf q$ lies on the manifold, then the correlation
 function $c_{\bf q}(\bbox{\lambda}) = \left<{\bf s}(\bbox{\lambda}),
{\bf q}\right>$  will reach its maximum value of unity in
${\cal S}_{\bf q}(\eta)$ when ${\bf q} = {\bf s}(\bbox{\lambda})$
and will fall off in all directions.  This means that the signals
 in the region which are further away from the filter are less
likely to be picked up as compared to those in the immediate
neighbourhood of the filter $\bf q$.

We assume that a finite subset of the normalised signal manifold has been
chosen to act as filters
by some suitable algorithm \cite{SD91}, which taken together span the manifold.
The number of filters will be determined by the available computing power.
Consider one of these filters $\bf q$, the region corresponding to it for a
threshold of $\eta$,
 ${\cal S}_{\bf q}(\eta)$, and an arbitrary normalised vector  ${\bf q}_o$
which belongs to $\cal V$
 but not necessarily $\cal S$. By correlating the vector ${\bf q}_o$ with
vectors in
${\cal S}_{\bf q}(\eta)$ we obtain the correlation function $c_{{\bf
q}_o}(\bbox{\lambda}) =
\left<{\bf s}(\bbox{\lambda}),{\bf q}_o\right>$.  We select that ${\bf q}_o$ to
serve as a more
optimal filter which maximises the average of this correlation function:
\begin{equation}
\left<{\bf s}(\bbox{\lambda}),{\bf q}_o\right>_{av} =
\frac{1}{\int\limits_{{\cal S}_{\bf q}(\eta)}
\sqrt{g}d^p\lambda}\int\limits_{{\cal S}_{\bf q}(\eta)}\left<{\bf
s}(\bbox{\lambda}),
{\bf q}_o\right>\sqrt{g}d^p\lambda= \left<
\int\limits_{{\cal S}_{\bf q}(\eta)}{\bf s}(\bbox{\lambda})d^p\lambda\ ,\ {\bf
q}_o\right>,
\end{equation}
where $g = det\left[g_{\mu\nu}\right]$. In the last step above
the  integration and the scalar product operations in the
above equation have been interchanged. Moreover, for the normalised
chirp manifold the metric does not depend upon the parameters in the coordinate
system we have
chosen and therefore, $g_{\mu\nu}$ is a constant and the factor $\sqrt{g}$
cancels.
We now use Schwarz's inequality to maximise the average correlation to obtain,
\begin{equation}
\label{avcorr}
{\bf q}_o = {\cal N}\int\limits_{{\cal S}_{\bf q}(\eta)}{\bf
s}(\bbox{\lambda})d^p\lambda,
\end{equation}
where $\cal N$ is just a normalisation constant.

	We implemented the above algorithm for filter placement for the case of
Newtonian
signals with certain modifications.
The normalised chirp waveform consists of three parameters $(\Phi,t_a,\tau_n)$.
If we keep $t_a$ and $\tau_n$ fixed then the tip of the signal vector traces
out a circle
as we vary $\Phi$. As any circle lies on a plane we can express a signal vector
as a
linear sum of two vectors where the two vectors differ only in the phase
parameter
and we take this phase difference to be $\pi/2$. Thus we need
only two mutually orthogonal
filters to span the phase parameter. The time of arrival parameter
$t_a$ is also a `convenient parameter' as by the use of fast Fourier transforms
the correlations
for arbitrary time of arrivals can be performed at one go.  It is therefore not
profitable for us
to maximise the average correlation over the phase parameter $\Phi$ and the
time of arrival
$t_a$.

	In view of the above restrictions we modified the the filter placement
algorithm.
We consider the correlation function for the case when the filter vector is on
the manifold.
We define the `line of curvature' to be the curve on the manifold along which
the
correlation function falls the least. Figure (\ref{fig_dets}) illustrates the
correlation function plotted
as a function of $\tau_0$ along the line of curvature.
It is seen from the contour diagram of the numerically computed
correlation function that the line of curvature lies nearly on the
submanifold $t_a + \tau_n = $ constant, of the normalised chirp manifold. We
take two curves
passing through the point $\bf q$ in the region ${\cal S}_{\bf q}(\eta)$:
\begin{enumerate}
\item 	$t_a + \tau_n = $ constant,\ \ $\Phi = 0$ and
\item	$t_a + \tau_n = $ constant,\ \ $\Phi = \pi/2$.
\end{enumerate}
We obtain one filter for each of the two curves by evaluating eq. \ref{avcorr}
where the
domains of integration correspond to the segments of the curves defined.

Having determined the two filters we again plot the correlation function along
the line of curvature as a function of $\tau_0$ in figure (\ref{fig_dets}).
The region of the manifold selected corresponds to a range of $5.8$~secs to
$6.0$~secs
in the parameter $\tau_0$. It is interesting to note that we could have
translated these
values keeping the difference same without affecting our results.
It can be seen that the correlation has a minimum at the center.
In order to get a flatter correlation curve we select a linear
combination of the original filter and the one obtained by averaging with
suitable
weights attached to each filter. This performs reasonably well as shown by the
thick curve in
the figure. The importance of having a flatter correlation function lies in the
fact
all the signals in a region can be picked up with equal efficiency and the drop
in the maximum possible correlation can be compensated for by lowering the
threshold.
The average correlation obtained for the optimal filter is only marginally
better than
that obtained for the filter placed on the manifold.

	In the discussion above, we had started with a fixed number of filters $n_F$
on the manifold and obtained another set of $n_F$ filters which performs
marginally
better than the former set.
Equivalently we can try to increase the span of each filter
retaining the same threshold  but reducing the number of filters
required. In Figure \ref{fig_dets} we observe that the optimal filter chosen
spans
the entire region considered with a threshold greater than $0.9$, whereas the
filter on the manifold
spans about half the region at the same threshold. This indicates that by
moving the filters out of the
manifold in the above manner it may be possible to reduce the number of filters
by a factor of two
or so. One must however, bear in mind that the bank of filters obtained in this
way are not optimal. There is scope to improve the scheme further. Our analysis
is indicative of this feature.

\subsection {Effective dimensionality of the parameter space of a second
order post-Newtonian waveform}
\label{2pnsec}

It has already been shown that the first post-Newtonian waveform is essentially
one-dimensional \cite{Sat94}.
We argue in this subsection that even the second post-Newtonian waveform is
essentially
one-dimensional and a one-dimensional lattice suffices to filter the waveform.

A Newtonian waveform is characterised by a set of three parameters
consisting of the time of arrival, the phase of the signal at the
time of arrival and chirp mass (equivalently, Newtonian chirp time).
In this case, for the purpose of detection,
one essentially needs to employ a one-dimensional lattice
of filters corresponding to the chirp mass, the time of arrival
being taken care by the fast Fourier transform algorithm and the
phase being determined using a two-dimensional basis of orthogonal templates.
When post-Newtonian corrections are included in the phase of the
waveform the number of parameters increases apparently implying that
one needs to use a two-dimensional lattice of filters corresponding to,
say, the chirp and reduced masses (equivalently the Newtonian and
post-Newtonian
chirp times) which in turn means that the number of templates through which
the detector output needs to be filtered goes up by several orders of
magnitude.  One of us (BSS) has recently shown that for the purpose
of detection it is sufficient to use a one-dimensional lattice of filters
even after first post-Newtonian corrections are included in the phase of
the waveform and the relevant parameter here is the sum of the Newtonian
and post-Newtonian chirp times. What happens when corrections beyond
the first post-Newtonian order are incorporated in the phase of the waveform?

The coalescing binary waveform is now available up to second post-Newtonian
order \cite {BDI,BDIWW}. Blanchet et al. argue that the phase correction
due to the second order post-Newtonian (2PN) term induces an accumulated
difference of 10 cycles in a total of 16000. Consequently, it is
important to incorporate the 2PN terms in the templates.
When the 2PN terms are included it is useful to consider that the full
waveform is parameterised by three additional parameters, corresponding to the
chirp times at the 1, 1.5 and 2PN order (cf. Sec \ref{waveforms}), as compared
to the Newtonian waveform. Of course, as far as the detection problem
is concerned there is only one additional parameter since the chirp times
are all functions of the two masses of the binary. However, for the purpose
of testing general relativity one can consider each of the chirp times
to be independent of the rest \cite{BS95,BS94}. Our problem now is
to find the dimensionality of the parameter space of a 2PN waveform.
To this end we consider the {\it ambiguity function} $C({\bbox
{\lambda}}^{'},{\bbox {\lambda}})$
which is nothing but
the correlation function of two normalised waveforms one of whose parameters
(${\bbox {\lambda}}$) are varied  by holding the parameters of the other fixed
(${\bbox {\lambda}}^{'}$) :
\begin {equation}
C({\bbox {\lambda}}^{'}, {\bbox {\lambda}}) = \left < q({\bbox {\lambda}}^{'}),
q({\bbox {\lambda}}) \right >;
\ \  \left < q({\bbox {\lambda}}^{'}), q({\bbox {\lambda}}^{'}) \right > =
\left < q({\bbox {\lambda}}^{\phantom{'}}),({\bbox {\lambda}}) \right > = 1.
\end {equation}
It is useful to think of ${\bbox {\lambda}}^{'}$ to be the parameters of a
template and ${\bbox {\lambda}}$ to be that of a signal. With this
interpretation
the ambiguity function simply gives the span of a filter in the parameter
space.

The ambiguity function for the full waveform is a four-dimensional
surface since there are four independent parameters. To explore the
effective dimensionality of the parameter we consider the set of
parameters to
be $\{t_a, \Phi, m_1, m_2\},$ where $m_1$ and $m_2$ are the two masses
of the binary. We have shown the contours of the ambiguity function maximised
over
$t_a$ and $\Phi$ (since these two parameters do not explicitly
need a lattice of templates) in Fig. \ref{ambcont}.
The template at the centre of the plot corresponds to a binary
waveform with $m_1=m_2=1.4 M_\odot$ and the signal parameters are
varied over the entire astrophysically interesting range of masses:
$m_1, m_2 \in [1.4,10]M_\odot.$ From these figures we find that
the ambiguity function is almost a constant along a particular
line in the $m_1$--$m_2$ plane. This means that a template at the
centre of the grid spans a relatively large area of the parameter
space by obtaining a correlation very close to unity for all signals
whose masses lie  on the curve along which the ambiguity function
roughly remains a constant. It turns out that the equation of this
curve is given by
\begin {equation}
\tau_{0} + \tau_{1} - \tau_{1.5} + \tau_{2} = \mbox{const.}
\end {equation}
Let us suppose we begin with a two-dimensional lattice of filters
corresponding to a certain grid (albeit, nonuniform) laid in the
$m_1$--$m_2$ plane. Several templates of this set will have their
total chirp time the same. Now with the aid of just one template
out of all those
that have the same chirp time we can effectively span the region that
is collectively spanned by all such filters.
More precisely, we will not have an appreciable loss in the SNR in
replacing all templates of a given total chirp time by one of them.
Consequently, the signal manifold can be spanned by  a one-dimensional
lattice of templates.

\section {Estimation of parameters}
\label{sec_espar}
In this section we discuss the accuracy at which the various parameters
of a coalescing binary system of stars can be estimated. All our results
are for a single interferometer of the initial LIGO-type which has a
lower frequency cutoff at 40 Hz. At present it is beyond the computer
resources available to us to carry out a simulation for the advanced
LIGO. In the first
part of this section we briefly review the well known results obtained
for the variances and covariances in the estimation of parameters using
analytical methods. Analytical methods assume that the SNR is sufficiently
large (the so-called strong signal approximation) and implicitly use a
continuum of the parameter
space. In reality, however, these assumptions are not necessarily valid
and hence it is essential to substantiate the results obtained using
analytical means by performing numerical simulations.
In the second part of this section we present an exhaustive discussion
of the Monte Carlo simulations we have performed to compute the errors and
covariances of different parameters.  As we shall discuss
below the computation of errors using the covariance matrix is erroneous even
at an SNR of 10-20. Our estimation of 1-$\sigma$ uncertainty in the
various parameters, at low SNRs, is substantially larger than those computed
using
the covariance matrix. However, for high values of the SNR
($>25$--$30$) Monte Carlo estimation agrees with the analytical results.

\subsection {Covariance matrix} \label {anacovar}

In recent years a number of authors have addressed issues related
to the variances expected in the parameter estimation
\cite{BS95,KKS95,F92,FC93,CF94,BS94,Kr,KLM93,PW95,JKKT},
In the standard method of computing the variances in the estimation of
parameters one makes the assumption that the SNR is
so large that
with the aid of such an approximation one can first construct
the Fisher information matrix $\Gamma_{\mu\nu}$
and then take its inverse to obtain the covariance matrix
$C_{\mu\nu}.$
In the strong signal approximation the Fisher information matrix
and the covariance matrix are given by
\begin {equation}
g_{\mu\nu} = \Gamma_{\mu\nu} = \left < {\partial {\bf s} \over \partial
\lambda^\mu},
{\partial {\bf s} \over \partial \lambda^\nu} \right >;
\ \ \ C_{\mu\nu} = \Gamma^{-1}{\mu\nu}.
\label{gamma1}
\end {equation}
As we have seen before the Fisher information and consequently the covariance
matrix is block diagonal and hence there is no cross-talk,
implying vanishing of the covariances between the amplitude and
the other parameters. Consequently, we need not construct, for the
purpose of Weiner filtering, templates corresponding to different amplitudes.

The judicious choice of parameters has also allowed a very elegant expression
for the Fisher information matrix. It is particularly interesting to note
that the information matrix does not depend on the values of the various
parameters, except for the amplitude parameter, and hence is a constant as far
as the parameter space of the normalised waveforms is considered.

 For the purpose of numerical simulations
it is convenient to choose the set
$\lambda^\mu = \{{\cal A}, t_a, \Phi, \tau_0, \tau_1\}$
where $\cal A$ is the amplitude parameter, $t_a$ and $\Phi$ are the
time of arrival of the signal and its phase at the time of arrival,
and $\tau_0$ and $\tau_1$ are the Newtonian and the post-Newtonian
coalescence times.
For noise in realistic detectors,
such as LIGO, the elements of the Fisher information matrix cannot be
expressed in a closed form and, for the set of parameters employed,
it is not useful to explicitly write
down the covariance matrix in terms of the various integrals since the
errors and covariances do not have any dependence on the parameters.
We thus evaluate the information matrix numerically and then take
its inverse to obtain the covariance matrix. Instead of dealing with
the covariance matrix $C$ it more instructive to work with the matrix
of standard deviations and correlation coefficients
$D$ which is related to the former by
\begin {equation}
{D_{\mu\nu}} = \left \{  \begin{array} {ll}
\sqrt {C_{\mu\nu}}, & {\rm if} \ \mu=\nu\\
C_{\mu\nu}/(\sigma_\mu \sigma_\nu) & {\rm if}\ \mu \ne \nu,
\end {array}\right.
\label {varcorr}
\end {equation}
where $\sigma_\mu={D}_{\mu\mu}$ is the 1-$\sigma$ uncertainty in
the parameter $\lambda_\mu.$ The off-diagonal elements of $D$ take
on values in the range $[-1, 1]$ indicating
how two different parameters are correlated: For $\mu\ne\nu,$
${D}_{\mu\nu}=1$, indicates that the two are perfectly correlated,
${D}_{\mu\nu}=-1$ means that they are perfectly anticorrelated and
${D}_{\mu\nu}=0$ implies that they are uncorrelated.
Since the information matrix is block-diagonal,
the amplitude parameter is totally uncorrelated
with the rest and thus an error in the measurement of $\cal A$ will not reflect
itself as an error in the estimation of the other parameters and vice versa.
In contrast, as we shall see below,
Newtonian chirp time is strongly anticorrelated to
post-Newtonian chirp time, which implies that if in a given experiment
$\tau_0$ happens to be estimated larger than its true value then
it is more likely that $\tau_1$ will be estimated to be lower than
its actual value. Such correlations are useful as far as detection
is concerned since they tend to reduce the number of templates needed
in filtering a given signal.
On the other hand, strong correlations
increase the volume in the parameter space to which an event can
be associated at a given confidence level.
It seems to be in general true that a given set of parameters do not satisfy
the twin properties of
having small covariances and reducing the effective dimension of the manifold
for the purpose of filtering.
We elaborate on this point below.

Given a region in a parameter space, it is useful to know the
proper volume (as defined by the metric) of the manifold corresponding
to the said region. In choosing a discrete set of filters for the detection
problem one has
to decide upon the maximum allowable drop in the correlation due to the finite
spacing. Once
this is fixed, the number of filters can be determined from the total volume of
the manifold.
For the  detection problem it is beneficial to have a small volume whereas if
the waveform is parameterized in such a way such that the manifold
corresponding to it covers
a large volume, then one can determine the parameters to a greater accuracy. As
a simple
example let us consider a two-dimensional toy model: $\bbox{\lambda} =
\{\lambda_1,\lambda_2\}$.
We compare different signal manifolds each corresponding to a different
parameterizations  of
the waveform.
We assume that the covariance matrix and its inverse, the Fisher information
matrix, to be:
\begin{equation}
\label{volumes}
C_{\mu\nu} = \left(     \begin{tabular}{cc}
                                $\sigma_{11}$&$\sigma_{12}$\\
                                $\sigma_{12}$&$\sigma_{22}$
                       \end{tabular}
            \right)
	\mbox{\ \  \ and\ \ \ }
		\Gamma_{\mu\nu} =
			\left(     \begin{tabular}{cc}
                                $\gamma_{11}$&$\gamma_{12}$\\
                                $\gamma_{12}$&$\gamma_{22}$
                       \end{tabular}
            \right)  = \frac{1}{\left(\sigma_{11}\sigma_{22}
-\sigma_{12}^2\right)}
	\left(
		\begin{tabular}{cc}
                               $\sigma_{22}$&$-\sigma_{12}$\\
                               $-\sigma_{12}$&$\sigma_{11}$
                \end{tabular}
        \right).
\end{equation}
The volume of the manifold corresponding to a region $\cal K$ of the parameter
space is given as,
\begin{equation}
\label{vol2}
V_{\cal K} = \int_{\cal K}\gamma_{11}\gamma_{22}
\left[1 -
\frac{\gamma_{12}^2}{\gamma_{11}\gamma_{22}}\right]d\lambda_1d\lambda_2
=  \int_{\cal K}\gamma_{11}\gamma_{22}\left[1 -
\epsilon\right]d\lambda_1d\lambda_2,
\end{equation}
where $\epsilon=\gamma_{12}^2/(\gamma_{11} \gamma_{22})$ is the correlation
coefficient.
It can be clearly seen that if the correlation coefficient is small,
keeping the variances in the parameters, fixed then the volume of the manifold
is maximal.
Since the parameters $\tau_0$ and $\tau_1$ are highly anticorrelated the proper
volume corresponding
to the region reduces to zero showing that the effective dimensionality of the
manifold is less.

Though, in principle, the variances and covariances are independent
of the chirp time, in reality there arises an indirect dependence
since one terminates a template at a frequency $f=1/(6^{3/2}\pi M)$
(where $M$ is the total mass of the binary) corresponding to the plunge
radius at $a=6M.$ Therefore, larger mass binaries are tracked
over a smaller bandwidth so much so that there is less frequency
band to distinguish between two chirps of large, but different, total mass.
Consequently, at a given SNR the error in the estimation of chirp times is
larger for greater mass binaries. This is reflected by the fact that
the integrals in eq. (\ref {gamma2})
are somewhat sensitive to the value of the upper cutoff.
(This also explains why the errors in the estimation of the chirp
and reduced masses are larger for greater mass binaries \cite{FC93,CF94}.)
In the following we assume that the noise
power spectral density is that corresponding to the initial LIGO
 for which a fit has been provided by Finn and
Chernoff \cite{FC93}.
For an SNR of 10 the matrix $D$ is given by
\begin {equation}
{D_{\mu\nu}} = \left (  \begin{array} {rrrrr}
1.0  &     0 &     0 &      0 \\
     &  8.37 & 0.999 & -0.999 \\
     &       &  3.16 & -0.998 \\
     &       &       &   8.4 \\
\end {array} \right ),
\end {equation}
for the Newtonian signal, and by
\begin {equation}
{D_{\mu\nu}} = \left (  \begin{array} {rrrrr}
1.0  &     0 &     0 &      0 &      0 \\
     &  20.4 & 0.997 & -0.972 &  0.911 \\
     &       &  6.7  & -0.954 &  0.881  \\
     &       &       &   45.1 & -0.982 \\
     &       &       &        &  25.98 \\
\end {array} \right ),
\end {equation}
for the first post-Newtonian corrected signal.
While computing varianaces and covariances the
integrals in equation (\ref {gamma3}) are evaluated by
chosing a finite upper limit of 1 kHz.
In the above matrices the entries are arranged in the order
$\{{\cal A}, t_a, \Phi, \tau_0\}$ in the Newtonian case,
$\{{\cal A}, t_a, \Phi, \tau_0, \tau_1\}$ in the post-Newtonian case,
off diagonal elements are dimensionless correlation coefficients and,
where appropriate, diagonal elements are in ms.
The values quoted in the case of the Newtonian waveform are consistent
with those obtained using a different set of parameters by
Finn and Chernoff \cite {FC93}.
In Fig.~\ref {fdependence} we have plotted $\sigma$'s, at an SNR of 10,
as a function of the upper frequency cutoff $f_c$ for Newtonian and
post-Newtonian chirp times and the instant of coalescence $t_C.$
We see that $\sigma$ is larger for higher mass binaries but this is because
we have fixed the SNR. However, if we consider binaries of different
total masses, all located at the
same distance, then a more massive binary produces a stronger SNR
so that in reality it may be possible to determine its parameters more
accurately than that of a lighter binary.
In Fig.~\ref{truedependence} we have plotted
$\sigma$'s for binaries all located at the same distance
as a function of total mass. We fix one of the masses at a value of
$1.4~M_\odot$
and vary the other from $1.4~M_\odot$ to $10~M_\odot$.
In computing the $\sigma$'s plotted in this
figure we have terminated the waveform at the plunge orbit and
normalised the SNR of a $10 M_\odot$-$1.4 M_\odot$ binary system
to 10. As a function of $M$ the
uncertainties in $\tau_0$ and $\tau_1$ initially falloff since the increase in
the SNR for
larger mass binaries more than compensates for the drop in the upper frequency
cutoff. However, for $M$ larger than a certain $M_0$ the increase in SNR is not
good enough
to compensate for the drop in $f_c$ so much so that the uncertainties in
$\tau_0$ and $\tau_1$
increase beyond $M_0$. The parameter $t_a$, however, falls off monotonically.

\subsection {Monte Carlo estimation of parameters}

In this Section we present the first in a series of efforts to compute
the covariance matrix of errors through numerical simulations for
a coalescing binary waveform
at various post-Newtonian orders. Analytical computation of the
covariance matrix, as in the previous section, gives us an
idea of the covariances and variances but, as we shall see in this
Section, at low SNR's it grossly underestimates the errors.
Quite apart from the fact that the assumptions made in deriving
the covariance matrix might be invalid at low SNR's, in a realistic detection
and data analysis, other problems, such as discreteness of the
lattice of templates, finite sampling of the data, etc., do occur.
It therefore seems necessary to check the analytical calculations
using numerical simulations to gain further insight into the
accuracy at which physical parameters can be measured.
This Section is divided into several parts: In the first part we
highlight different aspects of the simulation, in the second part we briefly
discuss the choice of templates for the simulation, in the third we elaborate
on the Monte Carlo method that we have adopted to carry out our simulations
and in the fourth we discuss problems that arise in a numerical simulation.
The results of our study are discussed in the next Section.

\subsubsection {Parameters of the simulation}

Let $s(t)$ be a signal of strength
$\cal A$ characterised by a set of parameters $\hat{\bbox{\lambda}}$
\begin {equation}
s(t; \hat{\bbox{\lambda}}) = {\cal A} h(t;\hat{{\bbox {\lambda}}}),\ \
\left<h,h\right >=1.
\end {equation}
In data analysis problems one considers a discrete version
$\{s^k | k=0,\ldots,N-1\}$ of the waveform $s(t)$ sampled at
uniform intervals in $t:$
\begin {equation}
s^k \equiv s(k\Delta); \ \ \ k=0,\ldots,N-1,
\end {equation}
where $\Delta$ denotes the constant interval between consecutive samples
and $N$ is the total number of samples. The sampled output $x^k$ of the
detector consists of the samples of the noise plus the signal:
\begin {equation}
x^k = n^k + s^k.
\end {equation}
The {\it sampling rate,} $f_s=\Delta^{-1}$
(also referred to as the {\it sampling frequency})
is the number of samples per unit time interval. In a data analysis
problem the sampling frequency is determined by the signal bandwidth.
If $B$ is the signal bandwidth, i.e., if the Fourier transform of the
signal is only nonzero over a certain interval $B,$ then it is
sufficient to sample at a rate $f_s=2B.$ In our case there is a lower
limit in the frequency response of the detector since the detector
noise gets very large
below a seismic cutoff at about 10--40 Hz. As mentioned in the last Section
there is also an upper limit in frequency up to which a chirp signal is
tracked since one does not accurately know the waveform beyond
the last stable orbit of the binary. This
corresponds to gravitational wave frequency $f_c=1/(6^{3/2}\pi M).$
For a neutron star-neutron star (ns-ns) binary $f_c \sim 1525$~Hz while
for a ns-black hole (of 10 $M_\odot$) (ns-bh) binary $f_c \sim 375$~Hz.
Due to constraints arising out of limited computational power, we terminate
waveforms at $750$ Hz even when $f_c$ is larger than 1000 Hz. Such a shutoff
is not expected to cause any spurious results since, even in the case of
least massive binaries of ns-ns, which we consider in this study,
more than 99 \% of the `energy' is extracted by the time the
signal reaches 750 Hz.
We have carried out simulations with two types of upper cutoff:
\begin {enumerate}
\item one in which all templates, irrespective of their total mass,
are shutoff beyond 750 Hz.,
\item  a second in which the upper frequency cutoff is chosen to be $750$ Hz
or $f_c,$ whichever is lower.
\end {enumerate}
Consistent with these cutoffs the sampling rate is always taken to be $2$ kHz.
(We have carried out simulations with higher sampling
rates and found no particular advantage in doing so
nor did we find appreciable changes in our results.)

In all our simulations,
as in the previous section, we take the detector noise power spectral density
$S$ to be that corresponding to  initial LIGO
\cite {FC93}. For the purpose of simulations we need to generate noise
corresponding to such a power spectrum. This is achieved by the following
three steps:
\begin {enumerate}
\item generate Gaussian white noise $n'^k$ with zero mean and unit variance,
$$\overline {n'^k}=0,\ \ \overline{n'^k n'^l} = \delta^{kl}$$
where an overbar denotes average over an ensemble,
\item compute its Fourier transform
$$\tilde n'^k\equiv \frac{1}{\sqrt{N}}\sum_{l=0}^{N-1} n'^l \exp (2\pi i k l /
N)$$
and
\item multiply the Fourier components by the square root of the
power spectral density,
$$\tilde n^k = \sqrt {S^k} \tilde n'^k.$$
\end {enumerate}
The resultant random process has the requisite power spectrum.
In the above, the second step can be eliminated since the Fourier transform
of a Gaussian random process is again a Gaussian, but with a different
variance. In other words we generate the noise directly in the Fourier domain.
The simulated detector output, in the presence of a signal
$s^k,$ in the Fourier domain is given by
\begin {equation}
\tilde x^k = \tilde n^k + \tilde s^k
\end {equation}
where $\tilde s^k$ is the discrete Fourier transform of the signal.

\subsubsection {Choice of templates}

To filter a Newtonian signal we employ the set of parameters
$\{t_a, \Phi, \tau_0\}$ and to filter a post-Newtonian signal we
employ the set $\{t_a, \Phi, \tau_0, \tau_1\}.$ Templates need not
explicitly be constructed for the time of arrival since computation
of the scalar product in the Fourier domain (and the availability of
fast Fourier transform (FFT) algorithms) takes care of the time of arrival
in essentially one computation ($N\log_2 N$ operations as opposed to
$N^2$ operations, where $N$ is the number of data points). Moreover,
there exists a two-dimensional basis for the phase parameter which
allows the computation of the best correlation with the aid of just
two filters.  Consequently, the parameter space is
essentially one-dimensional in the case of Newtonian signals
and two-dimensional in the case of post-Newtonian signals. (However,
as shown in Section \ref {2pnsec}
it is to be noted that for the purpose of detection the effective
dimensionality of the parameter space, even with the inclusion
of second post-Newtonian corrections, is only one-dimensional.) We adopt the
method described in Sathyaprakash and Dhurandhar \cite {SD91}
to determine the templates needed for chirp times. As described
in \cite {SD91,Sat94} filters uniformly spaced in $\tau_0$ and $\tau_1$
covers the parameters space efficiently.

\subsubsection {Monte Carlo method} \label {mcmethod}

In order to compute variances and covariances numerically,
we employ the Monte Carlo method. The basic idea here is to mimic
detection and estimation on a computer by performing
a very large number of simulations so as to minimize the uncertainties
induced by noise fluctuations.
In our simulations we generate a number of detector outputs $\{x^k\}$ each
corresponding to a definite signal $s^k(\hat{{\bbox {\lambda}}})$ of a certain
strength,
but corresponding to different realisations
of the random process $\{n^k\}.$ Computation of the covariance matrix involves
filtering each of these detector outputs through an a priori chosen
set (or lattice) of templates $\{q(t; {{\rm _t}\bbox{\lambda}_k}) |
k=1,\ldots,n_f\},$
where $n_f$ denotes the number of templates. The templates of the lattice
each has a distinct set of values of the {\em test} parameters
${{\rm _t}\bbox{\lambda}_k}$ and together
they span a sufficiently large volume in the parameter space.
The simulated detector output is correlated with each member of
the lattice to obtain the corresponding filtered output $C ({{\rm
_t}\bbox{\lambda}}_k):$
\begin {equation}
C({{\rm _t}\bbox{\lambda}}_k) = \left < x, q ({{\rm _t}\bbox{\lambda}}_k)
\right >.
\end {equation}
For a given realisation of noise a particular template obtains the
largest correlation and its parameters are the {\it measured} values
${{\rm _m}{\bbox{\lambda}}}$ of the signal parameters. Thus, the measured
values of the
parameters are defined by
\begin {equation}
\max_k C({{\rm _t}\bbox{\lambda}}_k) = C({{\rm _m}{\bbox{\lambda}}}).
\end {equation}
The measured values, being specific to a particular realisation of noise,
are random variables. Their average provides an
estimation ${{\rm _e}{\bbox{\lambda}}}$ of the true parameter values and their
variance
is a measure of the error $\sigma_{\bbox {\lambda}}$ in the estimation:
\begin {equation}
{{\rm _e}{\bbox {\lambda}}} = \overline{{\rm _m}{\bbox {\lambda}}},\ \ \ \ \
\sigma_{\bbox {\lambda}}^2 = \overline{ \left ( { {{\rm _m}{\bbox {\lambda}}} }
-
\overline{ {{\rm _m}{\bbox {\lambda}}}} \right )^2},  \ \ \ \
 D^{\mu\nu} = \overline{ {{\rm _m}\lambda}^\mu \ {{\rm _m}\lambda}^\nu
\over \sigma_\mu \sigma_\nu}, \ \ (\mu\ne \nu),
\label {mccovar1}
\end {equation}
where $D_{\mu\nu}$ are the correlation coefficients between
parameters $\lambda_\mu$ and $\lambda_\nu.$
In order to accurately determine $\sigma_{\bbox {\lambda}}$ a large number of
simulations
would be needed. If the measured values ${{\rm _m}{\bbox {\lambda}}}$ obey
Gaussian
statistics then after $N_{\rm t}$ trials the variance is determined
to a relative accuracy $1/\sqrt N_{\rm t}$ and estimated values can differ from
their
true values by $\sigma_{\bbox{\lambda}}/\sqrt N_{\rm t}.$ We have performed in
excess of 5000
trials, for each input signal, and thus our results are accurate to better
than 1 part in 70. Even more crucial than the number of simulations
is the number of templates used and their range in the parameter space.
We discuss these and other related issues next.

The actual templates chosen, say for the parameter $\tau_0,$
in a given `experiment' depend on the
true parameters of the signal, the number of noise realisations employed
and the expected value of the error. Let us suppose
we have a first guess of the error in $\tau_0$, say $\sigma_{\tau_0}$.
Then, we choose 51 uniformly spaced filters around $\hat {\tau}_0$ (where
$\hat{\tau_0}$ is the signal chirp time) such that:
\begin{equation}
{{\rm _t}\tau_0} \in \left [ \hat{\tau}_0 - 5 \sigma_{\tau_0},
\hat{\tau}_0 + 5 \sigma_{\tau_0} \right ].
\end{equation}
This implies that we are covering a 5$\sigma$ width in $\tau_0$ at
a resolution of
$\sigma_{\tau}/5.$ The probability, that a template between 4$\sigma$ and
5$\sigma$
from the true signal `clicks', being $\sim  6\times 10^{-5},$
we are on safe grounds since, in a given simulation, we consider no more
than 5000 trials.
(In comparison, the probability that a template between 3$\sigma$ and 4$\sigma$
clicks is $2.2\times 10^{-3}$ corresponding to an expected 13 events
in 5000 trials.) For a post-Newtonian signal, which in effect
needs to be spanned by a two-dimensional lattice of filters, the
above choice of templates implies a requirement of $2601 \times 2$
filters in all. Here a factor of 2 arises because for each filter
in the $\tau_0$--$\tau_1$ space we will need two templates corresponding
to the two independent values of the phase $\Phi$: $0$ and $\pi/2.$
In the case of a Newtonian signal, the lattice being one-dimensional,
one can afford a much higher resolution and range. Even with the aid of
a mere 201 templates we can probe at a $\sigma/10$ resolution
with a  $10\sigma$ range.

We start off a simulation with the pretension that there is
no knowledge of what the $\sigma_{{\bbox {\lambda}}}$'s are. Thus, we choose
as our first trial a very large $\sigma_{\bbox {\lambda}}$ and lay the lattice
of templates. With this lattice we perform a test run of 400 trials and
examine the distribution of the measured values. If the distribution
is not, as expected, a Gaussian then we alter $\sigma_{\bbox {\lambda}}$: we
decrease it if the distribution is too narrow and increase it if the
distribution is too wide and does not show the expected falloff.
In particular, we
make sure that the templates at the boundary of the chosen range do
not click even once and the skewness of the distribution
is negligible. When for a certain $\sigma_{\bbox {\lambda}}$ a
rough Gaussian distribution is observed then we carry out a
simulation with a larger number of trials (typically 5000).
We subject the measured values in this larger simulation to
the very same tests described above. We only consider for further
analysis such simulations which `pass' the above tests
and determine the estimates, variances and covariances of the
parameters using the measured values, with the aid of equation
(\ref {mccovar1}).

\subsubsection {Numerical errors and remedies}

There are several sources of numerical errors that tend to
bias the results of a simulation unless proper care is exercised
to rectify them. In this Section we point out the most
important ones and show how they can be taken care of.
Due to memory restrictions, the present version of our codes work with
single precision except the FFT, which is implemented in double precision.
In future implementations we plan to carry out all computations in
double precision. This will possibly reduce some of the numerical noise
that occurs, especially at high SNRs, in the present simulations.

\begin {enumerate}

\item {\it Orthonormality of filters:} For the sake of simplicity
it is essential that the filters are
normalised in the sense that their scalar product is
equal to unity: $ \left < q, q \right > = 1.  $
A waveform is normalized numerically using the discrete version of the scalar
product:
\begin {equation}
{\cal N} = {1 \over \sum_{k=0}^{N-1} S^{-1}_k \left | \tilde q^k \right |^2 }
\end {equation}

As mentioned earlier we use
a two-dimensional basis of filters for the phase parameter. Choosing
the two filters to be orthogonal to each other makes the maximisation
over phase easier. However, here care must be exercised.
Two filters $q(t; t_a, \tau_0, \tau_1, \Phi=0)$ and
$q(t; t_a, \tau_0, \tau_1, \Phi=\pi/2)$
are apparently orthogonal to each other. The numerically computed `angle'
between the two filters, chosen in this manner, often turns out to
be greater than $\sim 10^{-2}$ radians. Consequently, one obtains
erroneous correlations. In order to circumvent this problem we first
generate two filters that are roughly orthogonal to each other,
as above, and then use the Gram-Schmidt method
to orthogonalise the two vectors.
If an explicit numerical orthogonalisation such as this is {\em not}
implemented
then the measured values of the various parameters show spurious
oscillations in their distribution and the estimated values of the
parameters tend to get biased.

\item {\it Correlation function:}
The scalar product of two normalised templates
$q(t; {{\rm _t}{\bbox {\lambda}}})$ and  $q(t; {{\rm _t}{\bbox {\lambda}}}')$
is given by
\begin {equation}
C({{\rm _t}\lambda}_\mu, {{\rm _t}\lambda}_\nu') =
\left < q(t; {{\rm _t}\lambda}_\mu), q(t; {{\rm _t}\lambda}_\nu') \right >,
\ \ \left < q(t; {{\rm _t}{\bbox {\lambda}}}), q(t; {{\rm _t}{\bbox
{\lambda}}}) \right > =
\ \ \left < q(t; {{\rm _t}{\bbox {\lambda}}}'), q(t; {{\rm _t}{\bbox
{\lambda}}}') \right > = 1,
\end {equation}
where we have indicated the dependence of the scalar product
on the various parameters by explicitly writing down the parameter subscripts.
Let us fix the parameters of one of the templates, say ${{\rm _t}{\bbox
{\lambda}}},$
and vary the parameters of the other template. Of particular interest is the
behaviour of $C$ maximised over all but one of the parameters,
say $\lambda_{\nu_0}$:
\def\ooo #1 #2{\vphantom{S}^{\raise-0.5pt\hbox{$#1$}}_{\raise0.5pt
\hbox{$#2$}}}
\begin {equation}
C_{\rm max}({ {\rm _t}\lambda}_\mu, {{\rm _t} \lambda}_{\nu_0}') =
\max_{\ooo {\scriptstyle \ {{\rm _t}\lambda}_\nu'}
{\scriptstyle\nu \ne \nu_0}}
C ({{\rm _t}\lambda}_\mu, {{\rm _t} \lambda}_\nu').
\end {equation}
$C_{\rm max}$ is expected to drop monotonically as
$|\lambda_{\nu_0} - \lambda_{\nu_0}'|$ increases. However, we have
observed departures from such a behaviour possibly arising out of
numerical noise. Such a behaviour causes
bias in the estimation of parameters, and consequently in the determination of
their covariances, especially at high SNRs.  We have found no remedy to
this problem and some of our results at high SNRs may have biases
introduced by this effect. (Sampling the templates at a higher rate
did not help in curing this problem.)

\item {\it Grid effects:}
The parameters of a signal chosen for the purpose of
simulation and detection can in principle be anything and
in particular it need not correspond to any of the templates
of the lattice. However, in practice we find that whenever the signal
parameters do not correspond to a member of the lattice then the resultant
simulation has a bimodal distribution of the measured values. This is,
of course, expected since a signal not on the grid is picked up by
two nearest templates along each direction in the parameter space.
Sometimes we do find that the peaks corresponding to the bimodal distribution
does not belong to the nearest neighbour filters but slightly away.
This is related to the fact that the correlation function maximised over
the time of arrival and the phase of the signal falls off much too
slowly along the $\tau_0$--$\tau_1$ direction and small deviation
from a monotonic fall can cause biases. (Such biases would be
present in the case when a signal corresponds to one of the grid points
though the magnitude of the effect would be lower.) In order to
avoid this problem, and the consequent shifts in the estimation of
parameters and errors in the determination of variances and covariances,
we always choose the parameters of the signal to be that corresponding
to a template.

\item {\it Upper frequency cutoff and its effect on parameter estimation}

	The Fisher information matrix computed using the stationary phase
approximation
in Section \ref{sm} does
not include the effect of truncating the waveform at $a = 6M$ --- the plunge
cutoff.
 As mentioned before, we
have carried out simulations for both with, and without, incorporating the
upper cutoff.
As the covariance matrix incorporating the upper cutoff is not available we
have been able to compare the Monte Carlo results with the covariance matrix
only
for the latter case, where the cutoff is held fixed at $750$~Hz. If we
incorporate the
upper cutoff into the Monte Carlo simulations the errors in the parameters are
reduced drastically.
The effect of the upper cutoff is expected to be more important for the higher
mass binaries such
as the ones we have considered. The ambiguity function, in this case, no longer
remains independent
of the point on the manifold. In other words, the correlation between two
chirps depends not only on the difference between the parameters of the
signals, but also
on the absolute values of the parameters. The correlation surface also ceases
to be
symmetric {\em i.e.} the correlation between two chirps also depends on the
sign of
$\delta\lambda$, where, $\delta\lambda$ is the difference in the values of the
parameters. As the computational power
required for carrying out simulations for lower mass binaries is not available
to us
 the simulations have been restricted to ns-bh star binaries, where the effect
of the upper cutoff is important.

\item {\it Boundary effects}

	For the purpose of simulations a grid of filters has to be set up `around' the
signal.
The grid must be large  enough so that the estimated parameters do not
overshoot the boundary
of the grid. This causes a problem as every value in the \{$\tau_0,\tau_1$\}
plane
does not lead to a meaningful value for the masses of the binary system. This
does not
however prevent us to construct a waveform with such a value for
\{$\tau_0,\tau_1$\} even
though the signal in general does not correspond to any `real' binary system.
This is
valid, and even necessary, if we are to compare the numerical results with the
covariance matrix.

\item {\it Incorporating the cutoff in the presence of  boundary effects}

If we wish to incorporate the effects of the upper cutoff in simulations then
we run into a
serious problem, as we would have to know the total mass of the binary in order
to compute the
upper cutoff. For an arbitrary  \{$\tau_0,\tau_1$\} we can end up with negative
and even complex
values of the total mass and hence the upper cutoff at $a = 6M$ is not
meaningful.
Thus, we cannot even construct a waveform for an arbitrary combination of
\{$\tau_0,\tau_1$\}. Therefore, in such cases,
we restrict ourselves to simulations where the grid lies entirely
within valid limits for \{$\tau_0,\tau_1$\}.

\end {enumerate}

\subsection {Results and Discussion}

Our primary objective is to
measure the variances and covariances following the method described in
Sec. \ref{mcmethod} and study their departure from that predicted by
analytical means (cf. Sec. \ref {anacovar}).
We have carried out simulations for several values of the masses of
the binary and in each case the signal strength (which is a measure
of the SNR) is varied in the range 10--40. However,
since the variances and covariances are independent of the absolute
values of the parameters, for the parameter set that we employ,
results are only quoted corresponding to a typical binary system.
(See Sec. \ref {anacovar} for a discussion of the covariance matrix.)
Similar results are obtained in other cases too. We use two sets of
parameters to
describe our results. Monte Carlo simulations allow us to directly measure
the amplitude, the time of arrival, the phase at the time of arrival
and the chirp time(s). This is the set $\{{\cal A}, t_a, \Phi, \tau_0,
\tau_1\}.$

As we shall see below, the instant of coalescence can be measured much more
accurately than the time of arrival.
As a consequence of this, the direction to the source can be determined
at a {\em much greater accuracy} by employing $t_C$ as a
parameter instead of $t_a.$
Thus, we also quote estimates and errors for the parameter $t_C.$
Since the error in the estimation of the phase is quite large, even at
high SNRs, we ignore it in our discussions.

We first deal with the
Newtonian signal and highlight different aspects of the simulation
and discuss the results at length.  We then consider the
first post-Newtonian corrected signal.

\subsubsection {Newtonian signal}

In the case of Newtonian signals the parameter space is
effectively one-dimensional and, as mentioned earlier, in this case
the lattice of templates covers a $10$--$\sigma$ range of the parameters
at a resolution of $\sigma/10$ centred around the true parameters
of the signal.

In Fig. \ref {nmcfig} we have shown
the error $\sigma_{\bbox {\lambda}}$ in the estimation of parameters
$t_a,$ $\tau_0$ and $t_C,$
as a function of SNR, deduced using the covariance matrix as
solid lines and computed using Monte Carlo simulations as dotted lines.
The curve corresponding to the covariance matrix is obtained
using an upper frequency cutoff $f_c= 750$~Hz consistent with that
used in our simulations.  The errorbars in the estimation of
$\sigma_{{\bbox {\lambda}}}$'s are obtained using 4 simulations, each
with 4000 trials.  At low SNR's $\sigma_{{\bbox {\lambda}}}$'s have a larger
uncertainty, as expected, and for $\rho > 30$ this uncertainty is
negligible, and sometimes smaller than the thickness of the curves,
except in the case of $\sigma_{t_C}$ (see below, for a possible
explanation).

At low SNRs (10--15) there is a large
departure of the various $\sigma_{\bbox {\lambda}}$'s from that inferred using
the covariance matrix. At an SNR $\sim 17$ the two curves
merge (except in the case of $\sigma_{t_C}$)
indicating the validity of the covariance matrix results for
this and higher SNR's.  Interestingly,
the agreement between Monte Carlo simulation results and those
obtained using the covariance matrix is reached roughly at the same SNR
irrespective of the parameter in question.

We note that
in spite of the fact that the time of arrival and the chirp time have
large errors, the instant coalescence can be estimated very
accurately---an order of magnitude better than either.
What is puzzling, however, is that, in the case of $t_C,$ the Monte Carlo
curve drops below
the covariance matrix curve above an SNR of 15 and the two curves do
not seem to converge to one another
even at very high SNR's.  Coincident with the crossover of the two
curves, the error in the estimation of $\sigma_{t_C}$ increases,
contrary to what happens for the other parameters,
signalling that there is a large fluctuation in the estimation of
$\sigma_{t_C}.$ This behaviour, we guess, is an artifact of the low
value of the sampling rate. Of course, our sampling rate is sufficiently
high to respect the sampling theorem.  However, since $t_C$
is determined to an accuracy an order of
magnitude better than either $t_a$ or $\tau_0,$ a much higher
resolution in template-spacing would be needed for
determining the error in the instant of coalescence
than that used for estimating the errors in the
time of arrival or the chirp time(s).
Testing this claim, unfortunately, is beyond the computer resources at
our disposal since we would need a sampling rate of about 10 kHz with
a filter-spacing $10^{-4}$~s.  We hope to be able to resolve this issue
in course of time.
Nevertheless, the fact that the error in the estimation of $\sigma_{t_C}$
first decreases with SNR and increases only after the two curves
crossover, hints at the above possibility as a cause for this anomalous
behaviour. This effect is also observed in the case of a post-Newtonian
signal.

In Table \ref{nmctab}
we have given the actual signal parameters $\hat {{\bbox {\lambda}}},$
estimated
values of the parameters ${{\rm _e}{\bbox {\lambda}}}$ (cf. equation
(\ref{mccovar1}))
and the corresponding errors in their estimation $\sigma_{\bbox {\lambda}},$
for
several values of the SNR. Errors inferred from the covariance matrix
can be read off from Fig.~\ref{nmcfig}.
The estimated values are different from the true values, some of them being
overestimated
and some others being underestimated.
However, the deviations are often larger than what we expect.
In a simulation that uses $N_{\rm t}$ trials the estimated parameters
${{\rm _e}{\bbox {\lambda}}},$ assuming a Gaussian distribution for the
measured
parameters ${{\rm _m}{\bbox {\lambda}}},$ can be different from the true values
by
$\sigma_{\bbox {\lambda}}/\sqrt{N_{\rm t}}.$ (In contrast, the measured values
${{\rm _m}{\bbox {\lambda}}}$ can differ from their true values by
$\sigma_{\bbox {\lambda}}$
or more.) However, we often obtain a slightly larger deviation,
\begin {equation}
 2 {\sigma_{\bbox {\lambda}}\over \sqrt {N_{\rm t}}} \la \left |{{\rm_e}{\bbox
{\lambda}}} - \hat{{\bbox
{\lambda}}}\right | \la 3 {\sigma_{\bbox {\lambda}}\over \sqrt {N_{\rm t}}},
\end {equation}
and we are unable to resolve this discrepancy. A more concrete test
for the simulations is the histogram $n({{\rm _m}{\bbox {\lambda}}})$ of the
measured parameters, namely the frequency at which a given parameter
clicks in a simulation. This is shown plotted in Fig.~\ref {nhist}
for an SNR of 10.
The skewness
of the measured value is less than $10^{-2}.$ These results lend further
support to the Monte Carlo simulations. There are visible asymmetries
in the distributions of $\tau_0$ and $t_a$ and the asymmetries in the
two cases are of opposite sense. This can, of course, be understood from the
fact that $t_a$ and $\tau_0$ have a negative correlation coefficient.
The histogram of $t_C,$ even at an SNR of 10, has very few non-zero bins.
This of course reflects the fact that it is determined very accurately.
We are unable to resolve the central peak in $n(t_C)$ since,
as mentioned earlier, the sampling rate and resolution in
$\tau_0$ are not good enough to do so.

\subsubsection {Post-Newtonian signal}

As opposed to the Newtonian case
here we have essentially a two-dimensional lattice of filters corresponding
to $\tau_0$ and $\tau_1.$ For the purpose estimating variances and
covariances we lay a mesh consisting of $2601 \times 2$ uniformly spaced
filters around the true parameters of the signal.
As pointed out in Sec. \ref {detect} not all filters in the mesh, unlike
in the Newtonian case, would correspond to the waveform from a realistic
binary but that does not preclude their use in the Monte Carlo simulations.
We shall see that the results of our simulations lend
further support to the claim that for the purpose of detection,
the parameter space need only be one-dimensional \cite {Sat94}.
The results obtained for the first post-Newtonian
signal are qualitatively similar to that of a Newtonian signal
and we refer the reader, where appropriate, to the Newtonian case for
a more complete discussion.

In Fig. \ref {pnmcfig} we have shown
the error in the estimation of parameters $\tau_0,$ $\tau_1,$
$t_C$ and $t_a,$ clockwise from top left, respectively,
as a function of SNR.
The solid and dotted curves are as in Fig. \ref{nmcfig}.
Here again the upper frequency cutoff is taken to be 750 kHz.
Just as in the case of a Newtonian signal here too the results obtained from
Monte Carlo simulation are much higher than those obtained by employing
the covariance matrix. At an SNR of 10, which is the expected value
for a majority of events that initial LIGO and VIRGO interferometers
will observe, the Monte Carlo values are more than thrice as much as their
corresponding covariance matrix values and at an SNR of 15 the errors
are roughly twice that expected from the covariance matrix.
In absolute terms, however, the errors are still quite small compared to
the actual parameter values: for a ns-ns binary, at an SNR of 10,
\begin {equation}
{\sigma_{\tau_0}\over \tau_0} \sim 2.4 \%, \ \
{\sigma_{\tau_1}\over \tau_1} \sim 9.4 \%.
\end {equation}
At an SNR of 10 the time of arrival can be measured to an accuracy of
72 ms in contrast to a value of 20 ms expected from the covariance matrix.
As is well known, with the inclusion of the post-Newtonian terms error
in the estimation of the time of arrival and Newtonian chirp time
increases by about a factor of 2 and 3, respectively \cite{FC93,CF94}.

As in the Newtonian case here again we see that the Monte Carlo curves
approach the corresponding covariance matrix curves at a high SNR
the only difference being that the agreement is reached at a
higher SNR $\sim 25.$ For SNRs larger than this the
two curves are in perfect agreement with each other. As mentioned
earlier, $\sigma_{t_C}$ shows an anomalous behaviour possibly
arising out of insufficient resolution in the time of arrival and
the chirp times.

In Table \ref {pntab} we have listed the true parameters $\hat{{\bbox
{\lambda}}},$
the estimated values ${{\rm _e}{\bbox {\lambda}}}$ and the Monte Carlo errors
$\sigma_{\bbox {\lambda}}$ for different SNRs.
As in the Newtonian case here too the estimated values show a larger departure,
than expected,
from the true values. Histograms of the various measured parameters
including $t_C$ are shown in Fig. \ref {pnhist} for a signal strength of 10.
The skewness is below its standard deviation
of $\sqrt {15/N_{\rm t}}$ \cite {NUM}
indicating the Gaussian nature  of the various distributions. Even in the case
of a post-Newtonian signal $\sigma_{t_C}$ is so small that we only have three
non-zero bins in $n(t_C).$

We now turn attention to other, more general, issues arising out of the
simulations.

In Sec. \ref {2pnsec} we have argued, on the basis of the behaviour
of the noise-free correlation function, that the effective
dimensionality of the parameter space for the purpose of detection,
even in the case of a post-Newtonian signal, is only one-dimensional.
The results of our Monte Carlo simulation unambiguously show that this
is indeed true even in the presence of noise.
We investigated the two-dimensional histogram, that gives the number of
occurrances of
different templates in the lattice, in a particular simulation.
The templates that `click' are all aligned along the line
$\tau_0 + \tau_1=$~const. In a total of 5000 realisations corresponding
to this simulation there is only one instance when a filter outside
this region clicks giving a probability of less than $10^{-3}$ for
a template outside this region to give a maximum. Consequently, it is
only necessary to choose a single filter along this curve.

The distribution of the maximum correlation
$ C_{\rm max}(\hat {{\bbox{\lambda}}}, { {\rm _t}{\bbox {\lambda}}})$ obtained
from
different noise realisations needs a special mention since it
has an inherent bias.
In Fig. \ref {snrhist} we have shown the distribution of the maximum
correlation taken from one of our simulations corresponding to an
SNR of 10.
Notice a slight shift of the distribution towards a higher
value and this cannot be accommodated within the expected fluctuation
in the mean. The measured value
of the standard deviation $\sigma_{\cal A}$ is $0.95.$ Since the number
of simulations is 5000 we expect that the signal strength should differ
from the true value of 10 by no more than $\sigma_{\cal A}/\sqrt {5000}=0.014.$
However, the mean
value is 10.26 giving a deviation of $0.26$ which is about 20 times larger
than that expected.
This occurs at all SNRs and for both Newtonian and post-Newtonian signals.
This of course does not mean there is bug in the way
we are computing the maximum correlation.
In the process of maximisation, values greater
than the signal strength are favoured and consequently the mean of the
maximum correlation shows a shift towards a higher value.
This suggests that that the maximum of the correlation is a biased estimator of
the
signal strength.
We find, consistent with the covariance matrix calculation, that
the amplitude parameter is uncorrelated with the rest of the parameters;
crosscorrealtion coefficients $D_0\mu,$ $\mu\ne 0,$ (cf. eq. \ref {varcorr})
inferred from our Monte Carlo simulations are less than $\sim 10^{-6}.$

Finally, it is of interest to note how the phase parameter $\Phi$
is correlated with the time of arrival. A plot of ${{\rm _m}\Phi}$
versus ${{\rm _m}t}_a$ is shown in Fig. \ref{taphi}.
We find that the measured values of the time of arrival and the phase are
such that $2\pi f_{0} \ {{\rm _m}t}_a = {{\rm _m}\Phi},$ where $f_{0}$
has a value of approximately $51$~Hz.  When the
time of arrival shifts by more than a cycle of the signal the phase jumps
by a factor $2 \pi$ leading to the points seen in the top-left
and bottom-right corner of the figure. This makes the estimation
of the phase and the error in its estimation pretty involved.

\subsubsection {Incorporating the effects of upper cutoff}

As mentioned before,
incorporating an upper cutoff at the onset of the plunge has a drastic effect
on
the estimation of parameters. The incorporation of the upper cutoff is
implemented
by stopping the waveform when the instantaneous frequency reaches the frequency
associated with the onset of the plunge or
$750$~Hz, whichever is lower. However,
due to computational constraints we have carried out the simulations only for
high
mass binaries and hence the upper cutoff plays an important role in all our
simulations.
It is to be noted that the discussion of the
ambiguity function in section \ref{2pnsec} is not valid when the effect of the
upper
cutoff is taken into account though for low mass binaries, such as ns-ns
binaries,
the results there will still
be valid. The further dependence of the signal waveform on the total mass of
the
system through the upper cutoff means that we can estimate the individual
masses
more exactly though the computational power is bound to increase.

        In order to carry out the simulations for the present case we selected
a $10 M_\odot-1.2 M_\odot$ binary system as this enables us to choose the
filter grid well
within valid limits of $\tau_0$ and $\tau_1$. The simulations were carried out
for various values of
SNR starting from 10.
The histograms of the estimated parameters at an SNR of 10 is
shown in fig. \ref{fig_phistu}.
At this SNR  the errors obtained are $\sigma_{\tau_0} = 39.3$ ms,
$\sigma_{\tau_1} = 22.4$ ms, $\sigma_{t_a} = 23.1$ ms and $\sigma_{t_C} = 0.6$
ms.
These can be compared with the values in table \ref{pntab} and we can see that
except for
the parameter $t_C$ the errors are substantially lesser when the upper cutoff
is incorporated
into
the waveform. It is necessary to recompute the covariance matrix, as emphasized
before,
including the effect of the upper cutoff in order to compare these numerically
obtained values
with the covariance matrix. In order to do this it is not enough to replace the
upper limit
in the integral in eq. (\ref{gamma3}) with the upper cutoff. The waveform now
depends on
the total mass of the system through the upper cutoff and this information has
to be
incorporated into the waveform.

We carried out the simulations for various SNRs for the same value of masses
quoted above.
In the absence of the estimates of the covariance matrix when the upper cutoff
is incorporated
we assume that at an SNR of 40 the Monte Carlo estimates are consistent with
those of the covariance matrix.
In Figure~\ref{fig_psigu} we illustrate the results of our simulations.
The points on the dotted line
are the values obtained through Monte Carlo estimates and the continuous line
is obtained by fitting a
$1/\rho$ dependence of the errors on the SNR {\em assuming } consistency at an
SNR of 40.
It is seen as in the
previous simulations that except for the parameter $t_C$ the other parameters
are fairly consistent with a
$1/\rho$ dependence of the errors on the SNR at an SNR greater than 25.

\section{Conclusions}
\label{sec_con}

	In this paper we have introduced the use of differential geometry in studying
signal analysis
and have addressed issues pertaining to optimal detection strategies
of the chirp waveform.  We have also carried out Monte Carlo simulations to
check how well the
covariance matrix estimates the errors in the parameters of the chirp waveform.
We summarize below our main results.
\begin{enumerate}

\item	We have developed the concept of a signal manifold as a subset of a
finite dimensional
vector space of detector outputs. Using the correlation between two signal
vectors as a scalar
product we have induced a metric on the signal manifold.
With this geometric picture  it is possible
to pose the question of optimal detection in a more general setting. We
suggest that
the set of template waveforms  for the detection of the chirp signal need not
correspond to
any point on the chirp manifold. We propose an algorithm to choose templates
off the signal manifold and show that the drop in the correlation due
to the discreteness of the set of templates is reduced.
This algorithm, though certainly not the best,  motivates the search of more
efficient templates. In addition,
the chirp manifold corresponding to the second post-Newtonian waveform
is shown to be effectively
one-dimensional. This has important implications for the computational
requirement for the on-line detection of the chirp signal.
The use of a convenient set of parameters of the chirp waveform for carrying
out numerical and analytical simulations is stressed.
These parameters are such that the metric components are independent of
the parameters which implies that the manifold is flat and the corresponding
`coordinate system' is Cartesian.
As the metric defined is nothing but the Fisher information matrix, the
covariance matrix, being the
inverse of the Fisher information matrix, is also independent of the
parameters.

\item	Monte Carlo simulations have been carried out for the case of the initial
LIGO
to find out whether the actual errors in
the estimation of parameters is consistent with the values predicted by the
covariance matrix.
Simulations have been carried out for both the Newtonian as well as the
post-Newtonian
waveforms. We have restricted ourselves to the case of high mass binary
systems, such as bh-ns binaries, where the computational requirement is
not very heavy since the length of the data train, in such cases,
works out to be less than $8$~s. Nevertheless, as has
been shown in this paper, the covariance matrix is independent of the
parameters identified by us when waveforms are terminated at a constant
upper cutoff irrespective of their masses. Consequently, our results will hold
good
for binary systems of arbitrary masses. We point out the major
problems that arise while performing a numerical simulation
and, where appropriate, we suggest how they may be taken care of.
In particular, the effect of incorporating the upper cutoff in the
frequency of the gravitational wave at
the onset of the plunge, which essentially depends on the total mass
of the binary, is extremely important for high mass binaries.
Since the covariance matrix with the inclusion of such a mass-dependent
upper cutoff is not available we have carried out most of our simulations
using a  constant  upper cutoff. This enables us to directly compare the
results of our Monte Carlo simulations with those of the
analytically computed covariance matrix. Since
for binaries with total mass less than $5 M_\odot$ the plunge
induced upper cutoff is larger than that induced by the detector noise
these effects can be ignored for such binaries.

The numerical experiments indicate that the covariance matrix  underestimates
the errors in the
determination of the parameters even at SNRs as high as 20. In the Newtonian
case the correlation
coefficient of the time of arrival $t_a$ and the Newtonian chirp time
$\tau_0$ is found to be very close to $-1,$ so much so
that even at a SNR of $7.5$,  the instant of coalescence $t_C = t_a + \tau_0$
remains practically a constant.
The error in the estimation of $\tau_0$ for the post-Newtonian waveform is
about four times the
error obtained in the case of the Newtonian waveform at the same SNR. This is
expected as the
first post-Newtonian correction to the waveform introduces a new parameter
$\tau_1$
(called the first post-Newtonian chirp time) which is highly
(anti)correlated with $\tau_0$.

For the post-Newtonian waveform at an SNR of $10$ the error
in $\tau_0$ is about $3$ times that predicted by the covariance matrix.
This corresponds to a factor of $2$ in the  chirp mass $\cal M.$
The distributions for the parameters
have been obtained and are seen to be unimodal distributions and
are slightly more sharper than a Gaussian.
When the plunge induced upper cutoff is incorporated into the
waveform the errors in the estimation of parameters
decrease by a factor of about $2.5$. The correlation coefficient
between $\tau_0$ and $\tau_1$ is also found to decrease, which is
consistent with our discussion in Section \ref{anacovar}.

The results obtained suggest that higher moments than that used in
obtained in computing the covariance matrix may be important
in the determination of the errors in parameter estimation.
In the geometric picture this amounts to taking into account curvature effects,
either
intrinsic or extrinsic.

\item	We suggest that $t_C$ is a more suitable
parameter to estimate the direction to the source than the time of
arrival.
The latter is a kinematical parameter that fixes the time at which
the gravitational wave frequency reaches the lower cutoff of
the detector while the parameter $t_C$
has the physical significance of being the instant of coalescence.
At an SNR of 10 the error in $t_a$ is too large (20 ms)
to deduce the direction to the source accurately, whereas the error
in the parameter $t_C$ is less than $0.5$~ms.
This will further
go down substantially for the advanced LIGO. A detailed analysis of the
determination of the direction to the binary using
delays in $t_C$ between different detectors
is carried out in Bhawal and Dhurandhar \cite{BD95}
(also see \cite {BSD95}).

\end{enumerate}

	We now suggest further work which needs to be done along the lines of this
paper.
A  full understanding of the chirp  signal manifold when higher post-Newtonian
corrections
are incorporated into the waveform is in order. This will help in the
development of more efficient algorithms
for the choice of templates in the detection problem and facilitate reduction
in computational time.
The Monte Carlo simulations which we have carried our
are for the case of a binary waveform correct up to first post-Newtonian order.
Moreover, only circular orbits are considered.
The effect of eccentricity is currently being investigated \cite{GBS95}.
 Performing simulations when higher post-Newtonian corrections
 are taken into account
calls for an immense amount of computational time. Fortunately,
matched filtering algorithm being amenable to parallelization
\cite {SD93}, one could aim
at using the massively parallel computers, that are now becoming available
the world over, in performing such simulations.

\acknowledgments
The authors would like to thank the members of the gravitational wave group at
IUCAA
especially S.D. Mohanty and B. Bhawal
for many helpful discussions. R.B. is being supported by the Senior Research
Fellowship
of CSIR, India.

\begin{table}
\label{nmctab}
\caption{The estimated value of the parameters and their errors
${\rm _e}{\lambda}$ $(\sigma_{{\lambda}})$
for the Newtonian waveform. The actual values of the parameters taken were
$\hat {\tau}_0 = 5558.0$ ms and
$\hat{t}_a = 200.0$ ms.  Except for the parameter ${\cal A}$ all values
are quoted in ms.}

\begin{tabular}{cccccc}
&$\rho=7.5$&$\rho=10.0$&$\rho=12.5$&$\rho=15.0$&$\rho=20.0$\\
\hline\\
${\rm _e}\tau_0\ (\sigma_{\tau_0})$&$5557.4\ (29.7)$&$5557.3\ (15.6)$&$5557.6\
(9.0)$&$5557.7\ (6.2)$&$5558.5\ (3.9)$\\
${\rm _e}t_a\ (\sigma_{t_a})$&$200.6\ (29.4)$& $200.7\ (15.5)$&$200.4\ (8.9)$&
$200.3\ (6.2)$& $199.5\ (3.9)$\\
${\rm _e}t_C\ (\sigma_{t_C})$&$5758.0\ (0.30)$& $5758.0\ (0.20)$& $5758.0\
(0.10)$&$5758.0\ (0.08)$& $5758\ (0.02)$\\
${\rm _e}{\cal A}\ (\sigma_{\cal A})$&$7.696\ (0.96)$&$10.12\ (0.98)$& $12.582\
(0.99)$&$15.067\ (0.99)$&$20.05\ (0.99)$\\
\end{tabular}
\end{table}

\begin{table}
\label{pntab}
\caption{The estimated value of the parameters and the errors in their
estimation
${\rm _e}{{\lambda}}$ $(\sigma_{{\lambda}}$)
for the post-Newtonian waveform.  The actual values of the parameters
taken were $\tau_0 = 5558.0$ ms, $\tau_1 = 684.0$ ms and
$t_a = 300.0$ ms  Except for the parameter $\cal A$ all values are quoted in
ms.}
\begin{tabular}{cccccc}
&$\rho=10.0$&$\rho=15.0$&$\rho=20.0$&$\rho=25.0$&$\rho=30.0$\\
\hline\\
${\rm _e}\tau_0\ (\sigma_{\tau_0})$ &$5554.9\ (136.1)$&$5555.9\
(64.8)$&$5557.2\ (30.1)$&$5556.7\ (19.3)$ &$5558.7\ (13.6)$\\
${\rm _e}\tau_1\ (\sigma_{\tau_1})$& $685.55\ (65)$ &$685.2\ (32.6)$&$684.6\
(16.4)$&$684.8\ (10.5)$&$683.6\ (7.2)$\\
${\rm _e}t_a\ (\sigma_{t_a})$&$301.49\ (73.26)$&$300.88\ (33.4)$ &$300.28\
(14.5)$ &$300.5\ (9.4)$&$299.7\ (6.8)$\\
${\rm _e}t_C\ (\sigma_{t_C})$ &$6.542\ (0.54)$&$6.542\ (0.30)$& $6.542\ (0.17)$
& $6.542\ (0.10)$&$6.542\ (0.06)$ \\
${\rm _e}{\cal A}\ (\sigma_{\cal A})$& $10.25\ (0.97)$&$15.15\ (0.98)$&$20.11\
(.99)$&$20.1\ (0.99)$& $30.1\ (0.99)$\\
\end{tabular}
\end{table}

\begin{figure}
\caption {This figure illustrates the correlation function along the line of
curvature as
a function of $\tau_0$, the Newtonian chirp time, for the following three
cases:
(i) when the filter is placed on the manifold (dotted line),
(ii) when the filter is the `average' signal vector over the region (dashed
line) and
(iii) when the filter is chosen to be an appropriate
linear combination of the previous two filters (solid line).
}
\label{fig_dets}
\end{figure}

\begin{figure}
\caption{Contour diagram of the ambiguity function for the second
post-Newtonian case.}
\label{ambcont}
\end{figure}

\begin{figure}
\caption{
Dependence of the errors in the estimation of the parameters
of the post-Newtonian waveform on the upper cutoff frequency.
The SNR is kept fixed at a value of 10.}
 \label{fdependence}
\end{figure}

\begin{figure}
\caption{Dependence of the errors in the estimation of the parameters
of the post-Newtonian waveform on the total mass of the binary keeping the
distance to the binary fixed.
The waveforms are cutoff at frequencies corresponding to the onset of the
plunge orbit
and the SNR is normalized at a value of 10 for a 1.4 $M_\odot$-10 $M_\odot$
binary.
}
\label{truedependence}
\end{figure}

\begin{figure}
\caption{Dependence of the errors in the estimation of parameters of the {\em
Newtonian} waveform
{\em i.e.} \{$\sigma_{\tau_0},\sigma_{t_a},\sigma_{t_C}$\}
as a function of SNR. The solid line
represents the analytically computed errors whereas the dotted line represents
the errors obtained
through Monte Carlo simulations.
}\label{nmcfig}
\end{figure}

\begin{figure}
\caption{Distributions of the measured values of the parameters for the case of
Newtonian
signal. The total number of noise realizations is 5000.}\label{nhist}
\end{figure}

\begin{figure}
\caption{Dependence of the errors in the estimation of parameters of the
post-Newtonian waveform
{\em i.e.} \{$\sigma_{\tau_0},\sigma_{\tau_1},\sigma_{t_a},\sigma_{t_C}$\}
as a function of SNR. The solid line
represents the analytically computed errors whereas the dotted line represents
the errors obtained
through Monte Carlo simulations.}
\label{pnmcfig}
\end{figure}

\begin{figure}
\caption{Distributions of the measured parameters of the post-Newtonian
waveform at a SNR of 25.
The number of noise realizations is 5000.
}\label{pnhist}
\end{figure}

\begin{figure}
\caption{The histogram of random variable $_{\rm m}{\cal A}$ at an SNR of
20. The variance in this parameter is
independent of the SNR and is approximately equal to unity.}\label{snrhist}
\end{figure}

\begin{figure}
\caption{The correlation between $t_a$ and $\Phi$ is illustrated. The phase
parameter simply
follows the time of arrival parameter. }\label{taphi}
\end{figure}

\begin{figure}
\caption{Distributions for the measured parameters of the post-Newtonian
waveform
incorporating the effect of the upper cutoff. The errors in the determination
of the parameters
is much smaller in this case. The total number of noise realizations is 5000.
 }\label{fig_phistu}
\end{figure}

\begin{figure}
\caption{The dependence of the errors on the SNR $(\rho)$
when the upper cutoff is incorporated into the waveform}
\label{fig_psigu}
\end{figure}

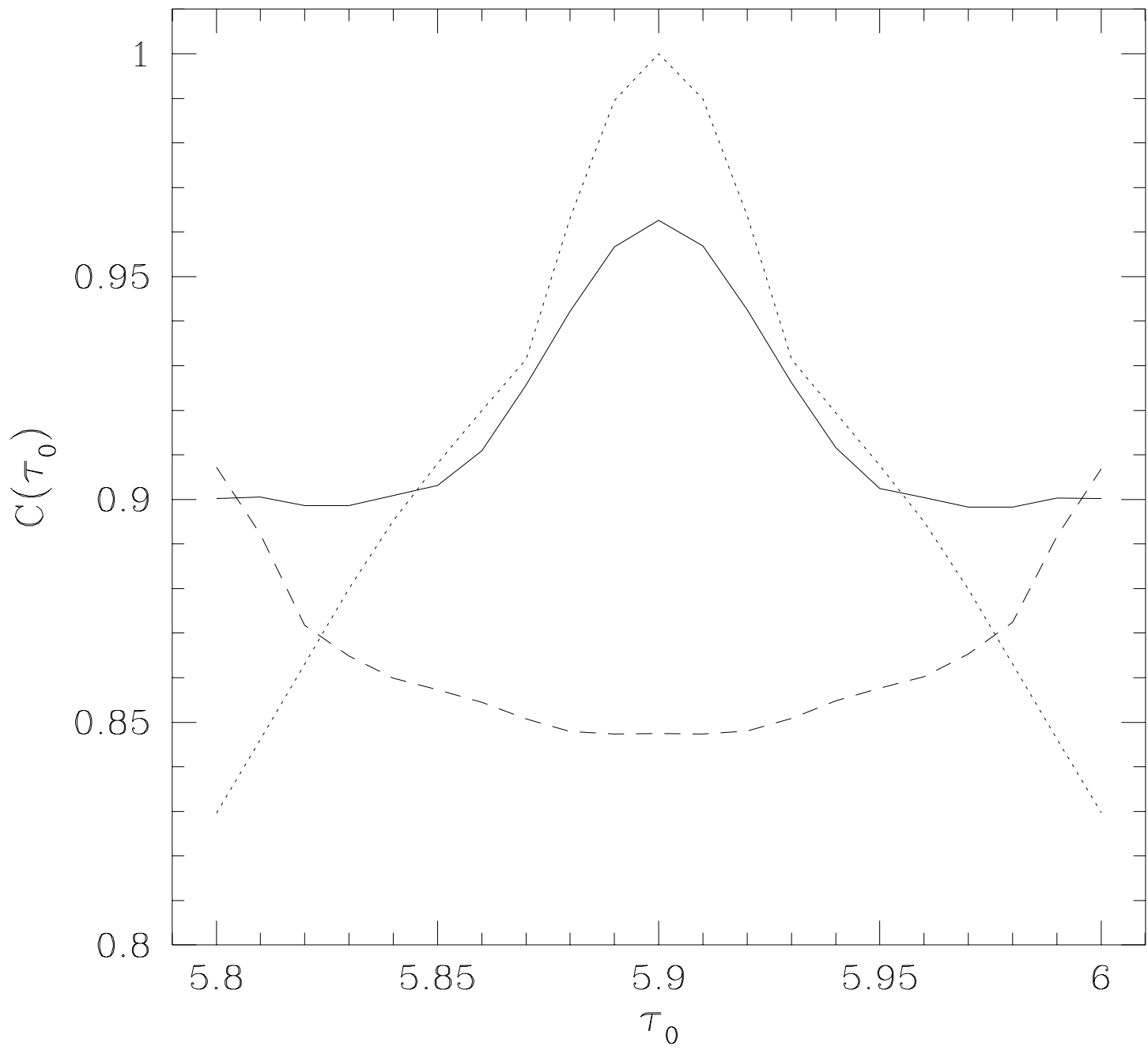
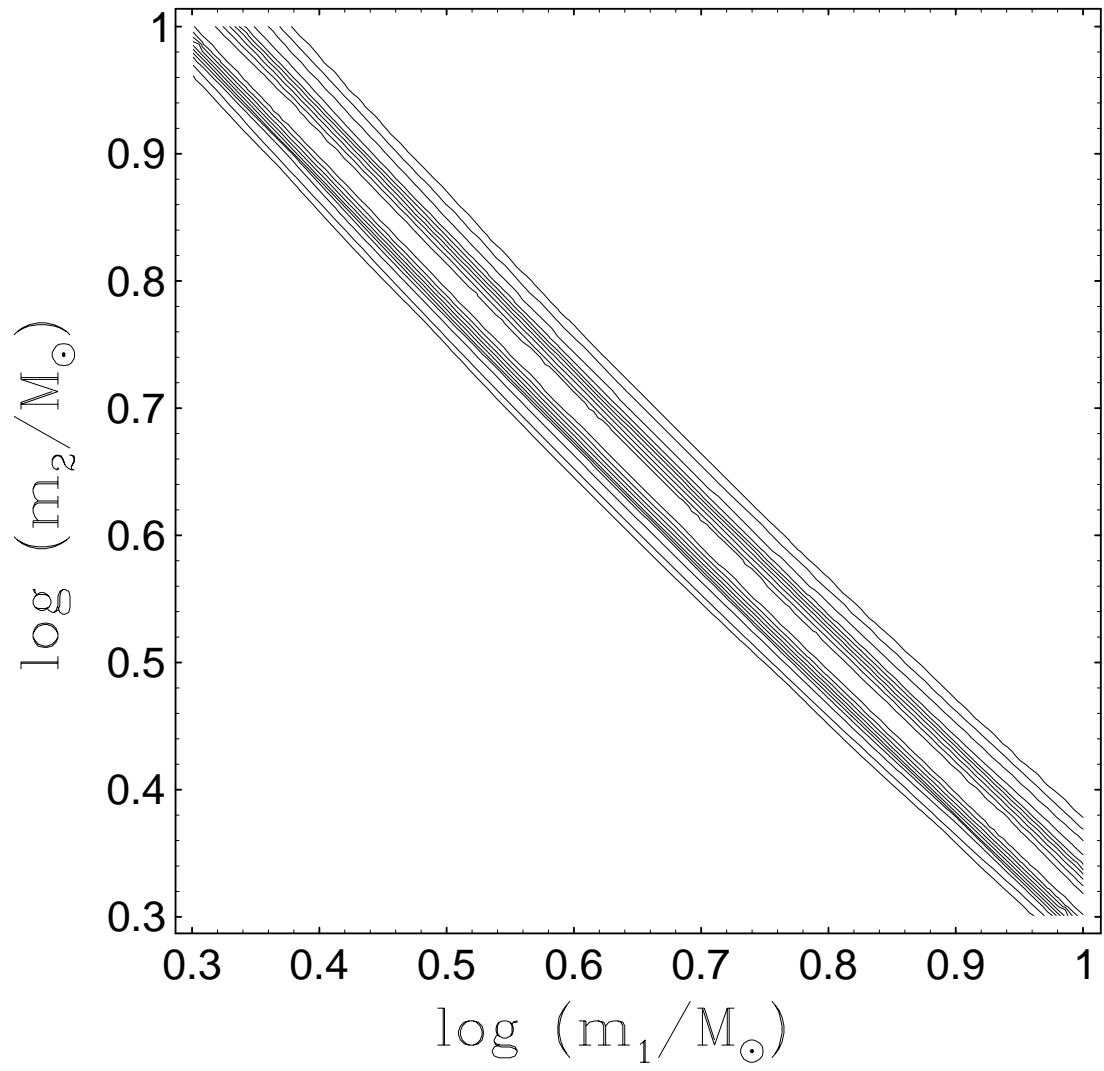
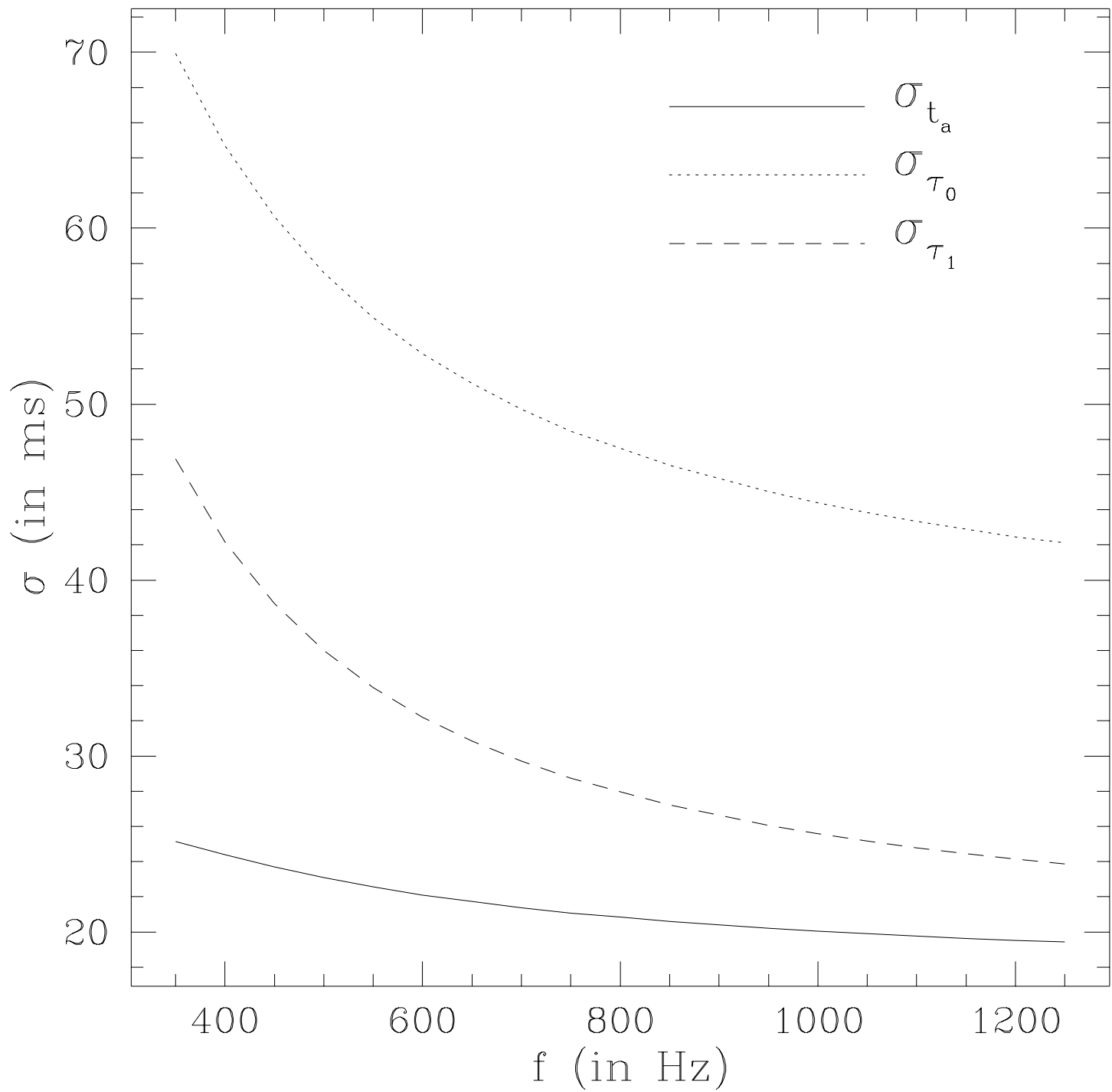
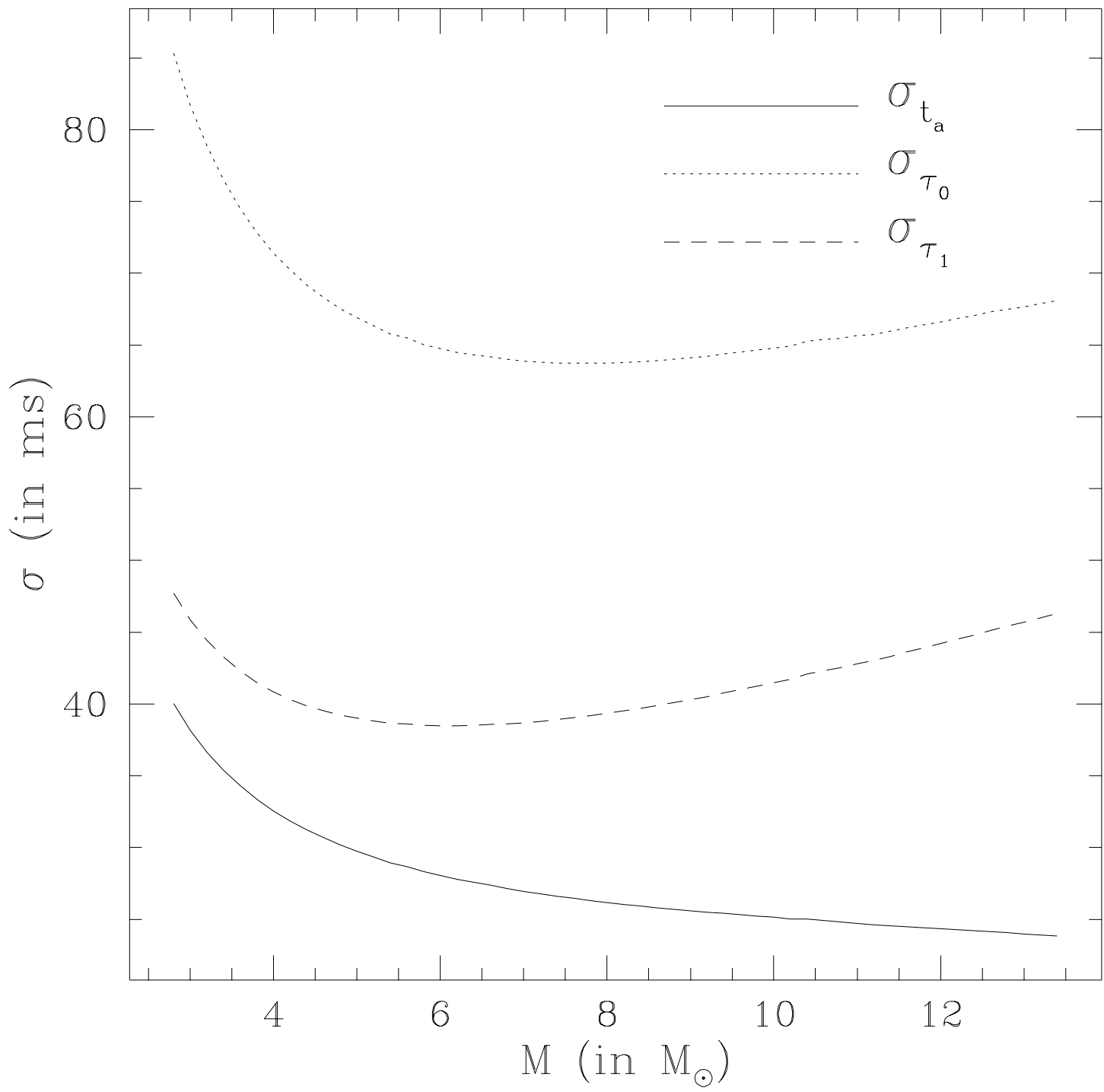
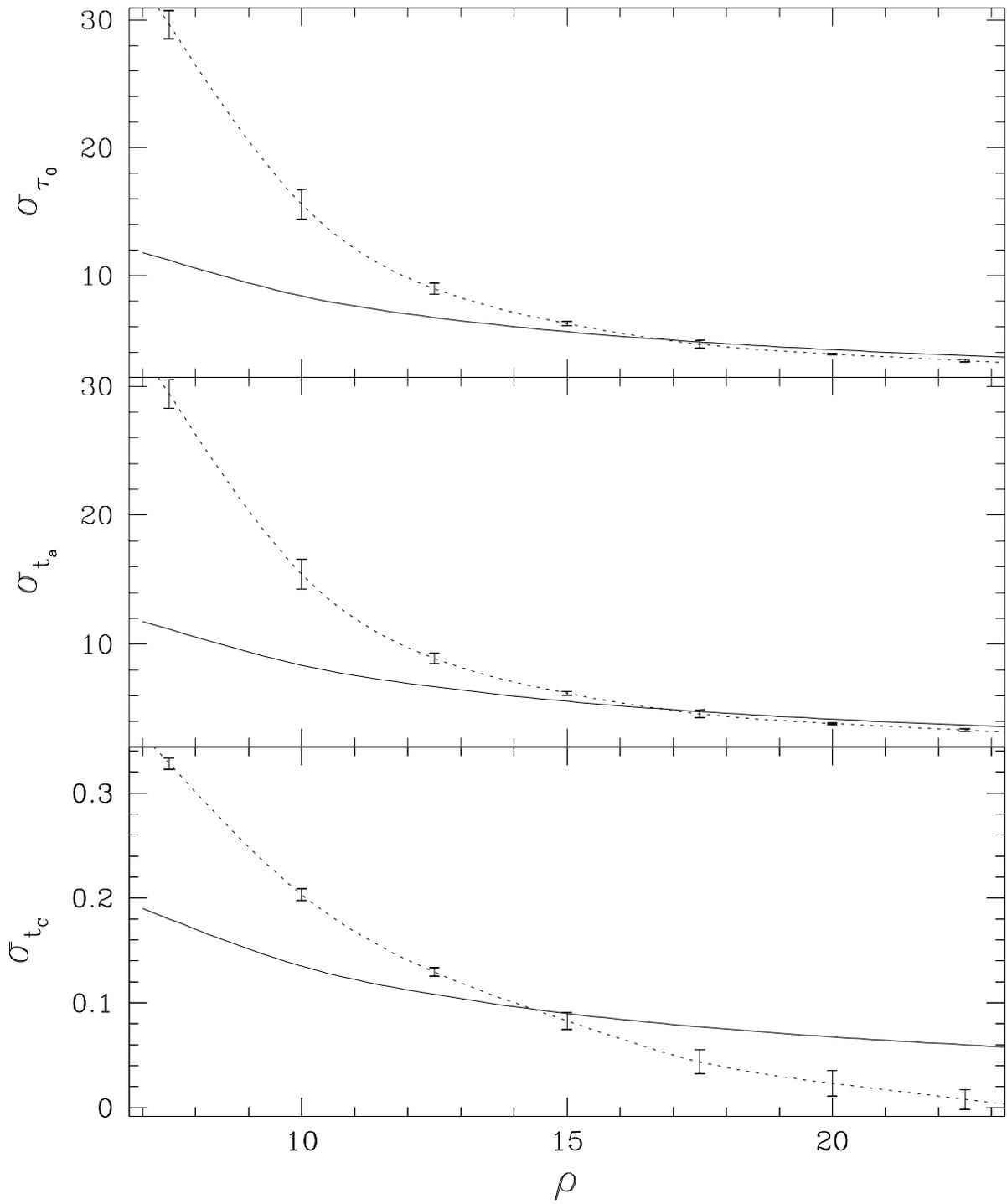
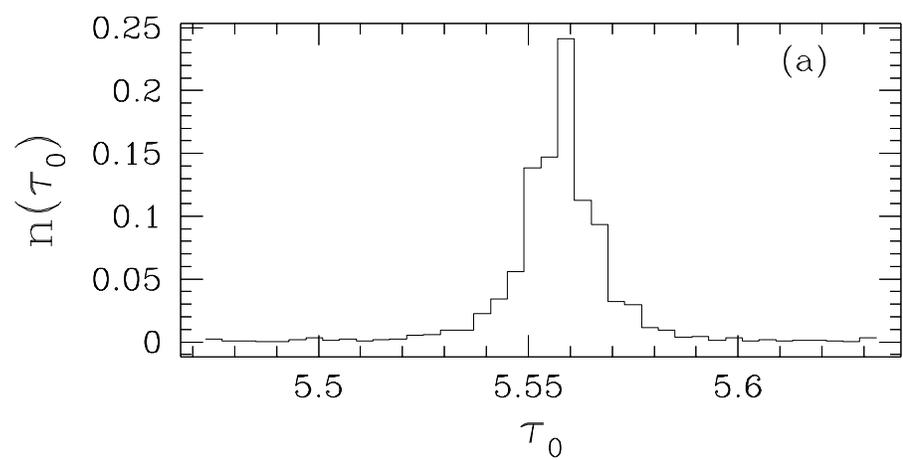
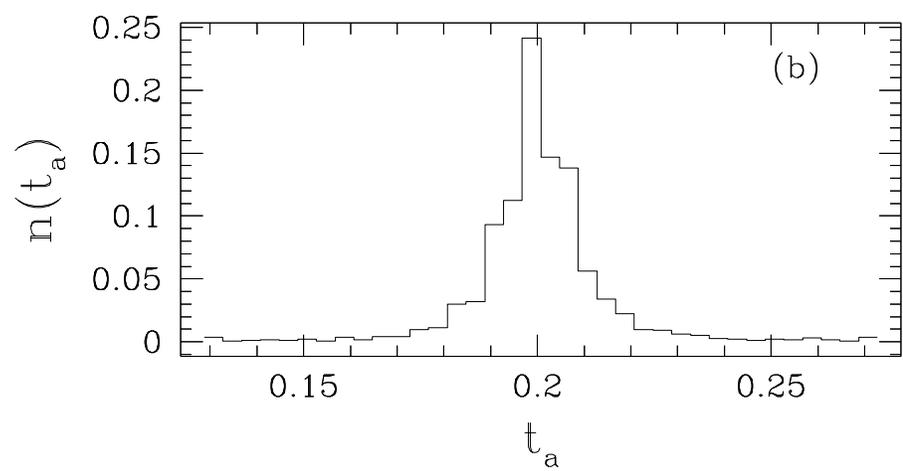
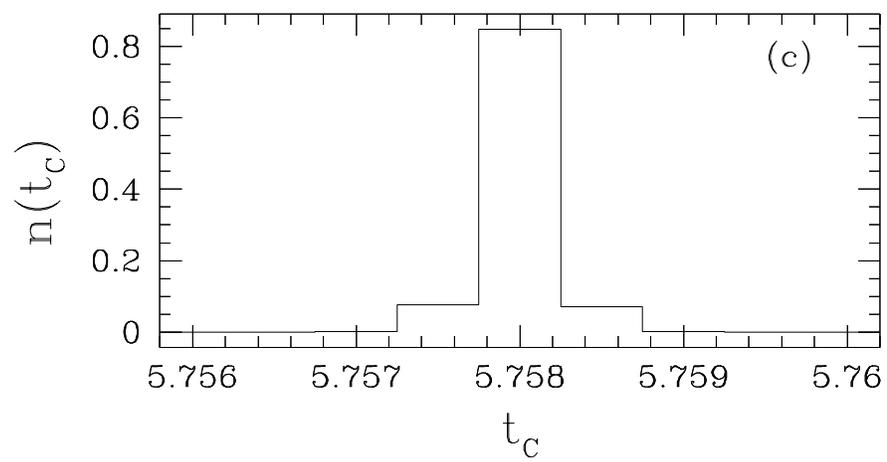
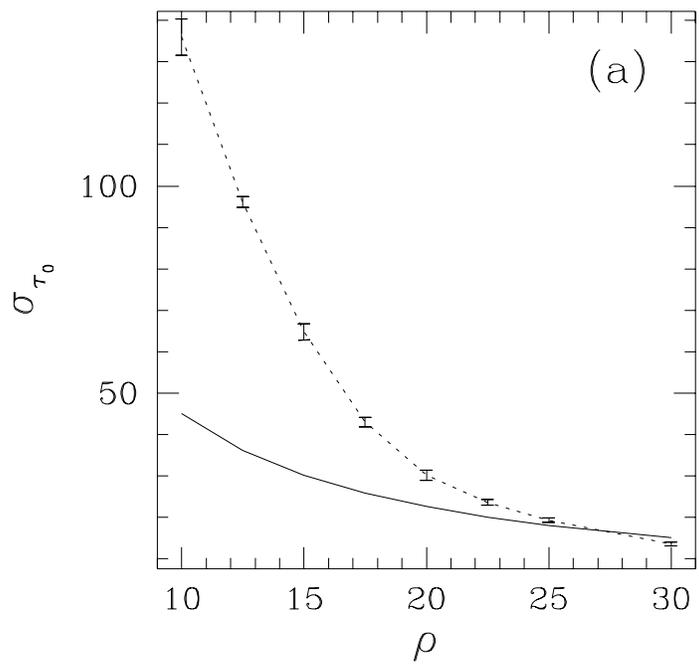
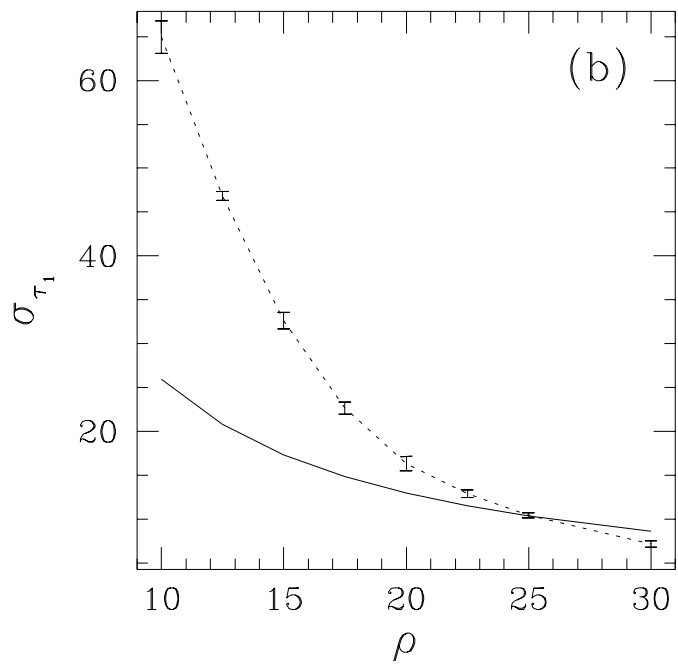
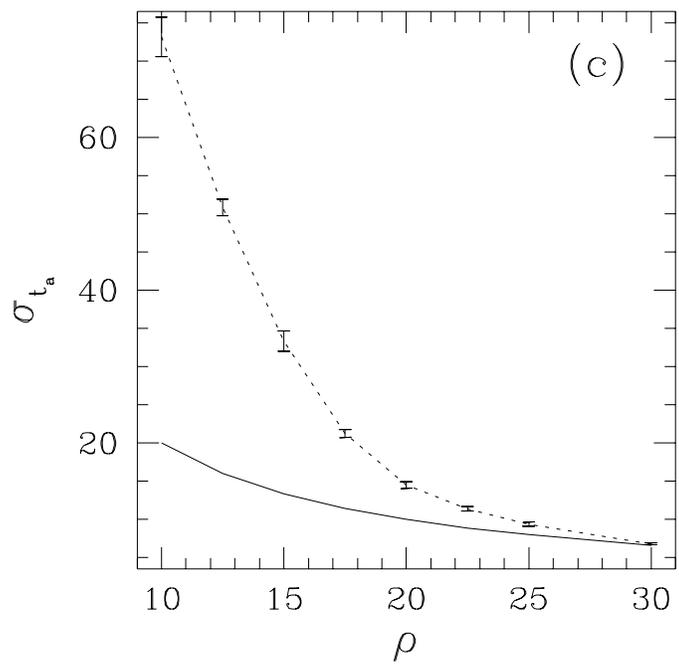
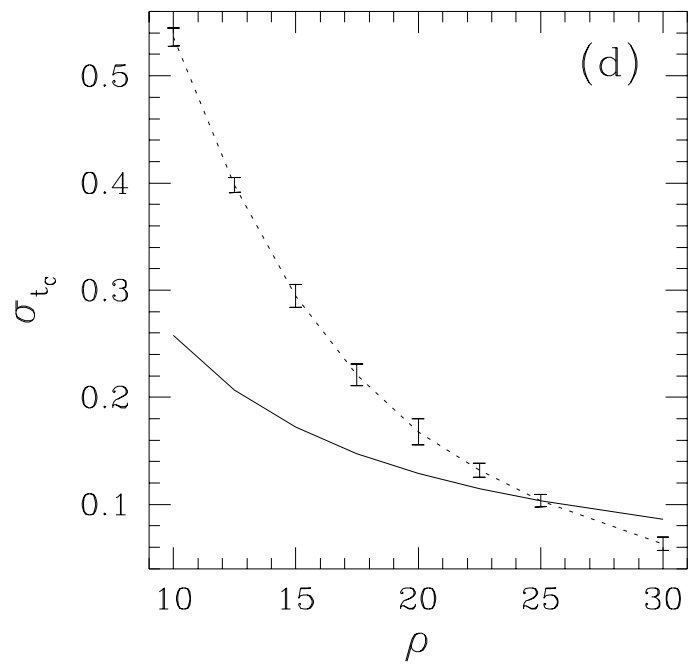
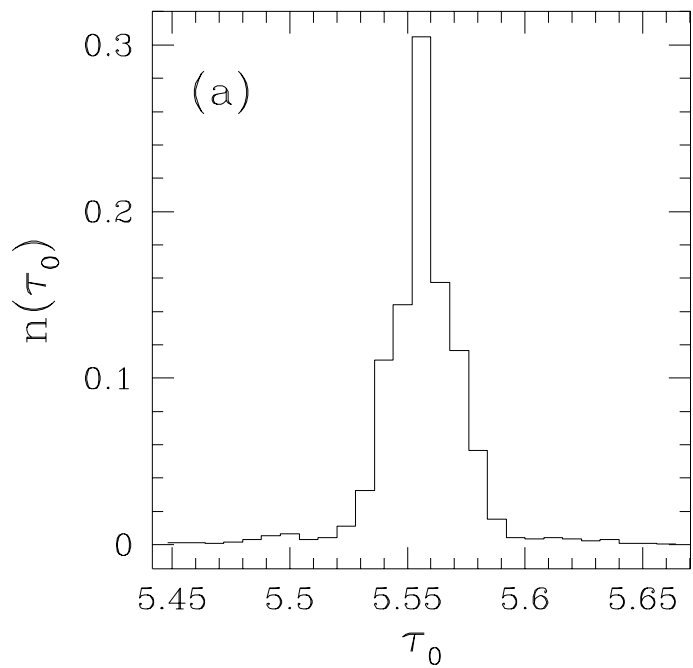
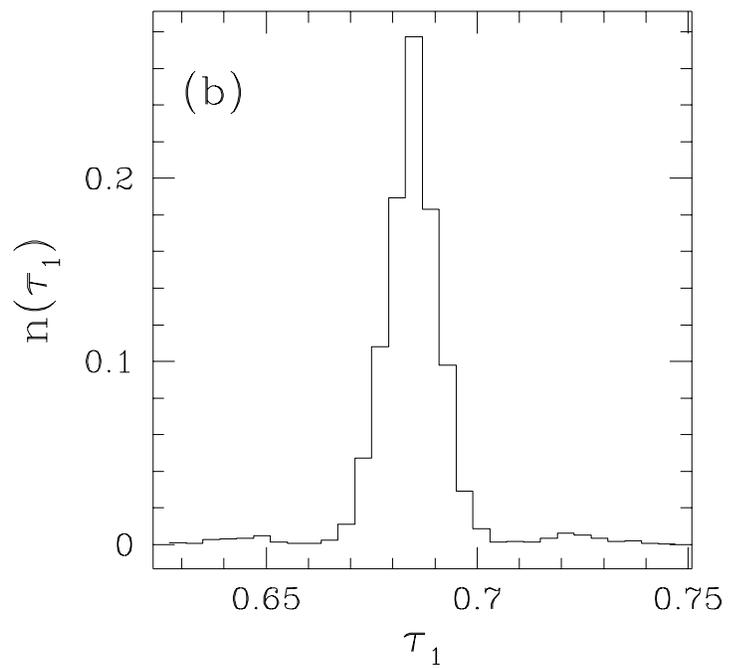
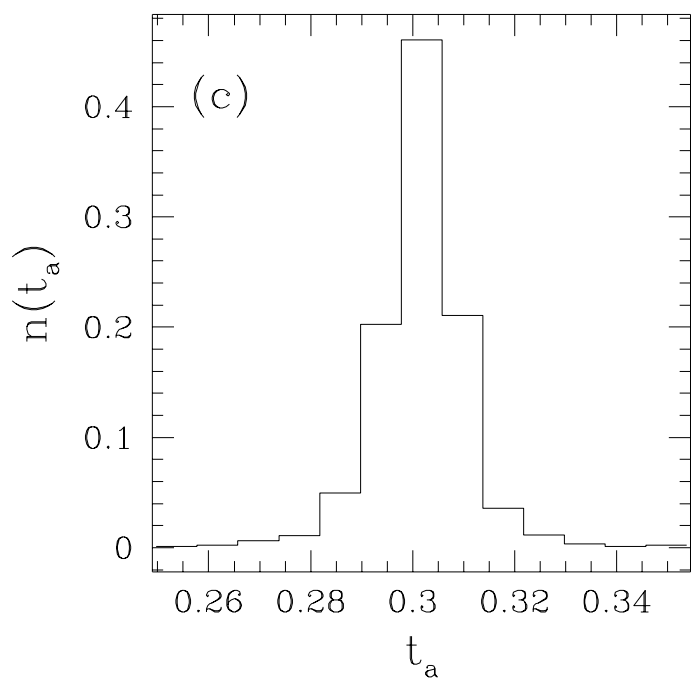
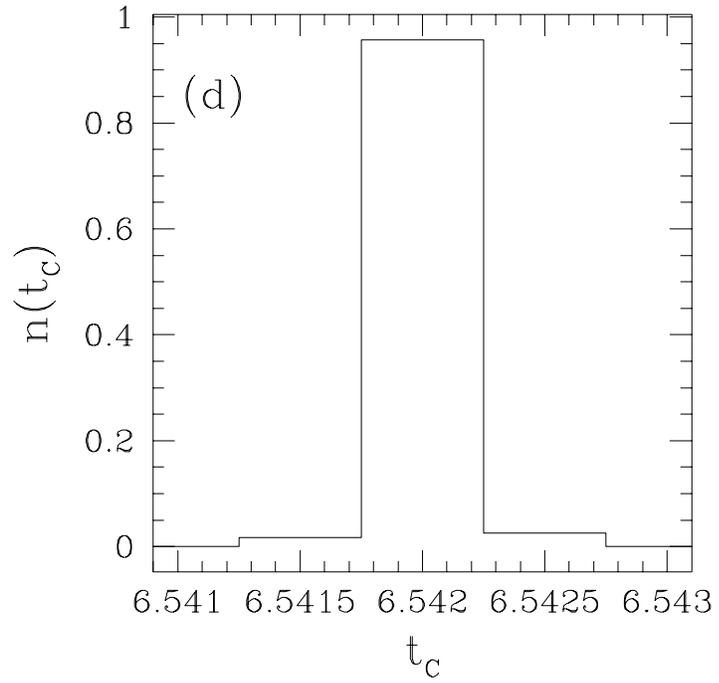
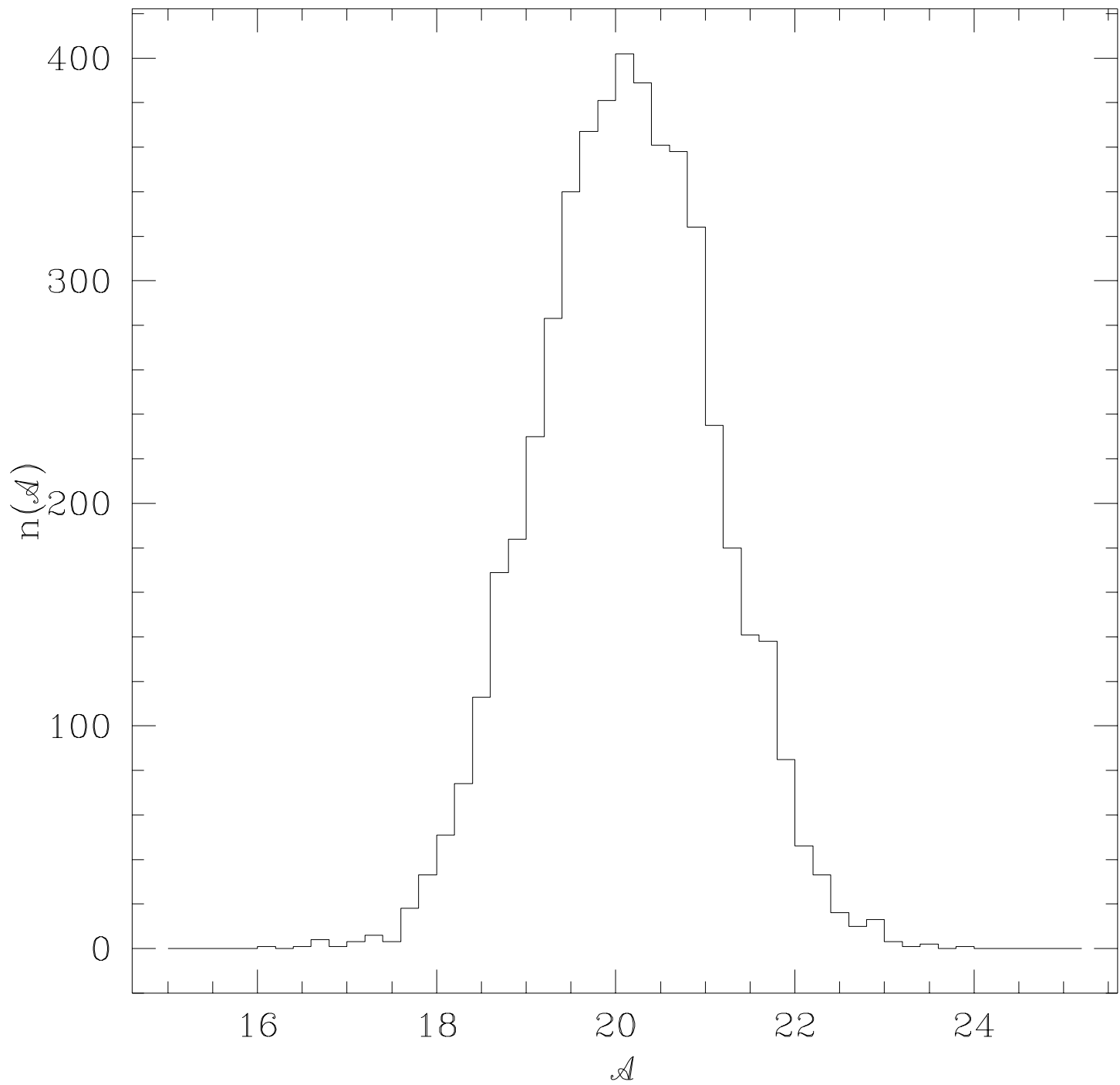
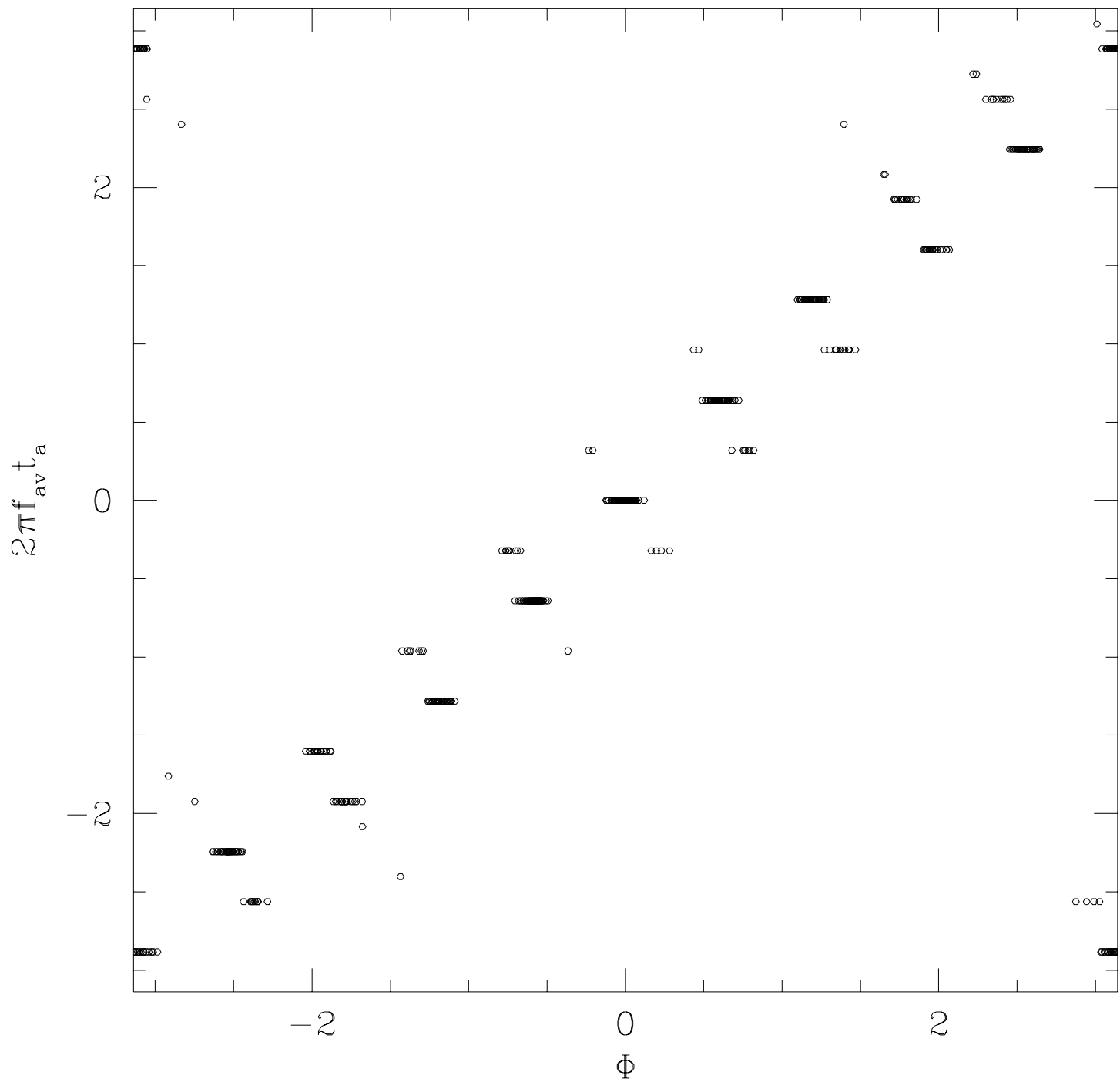
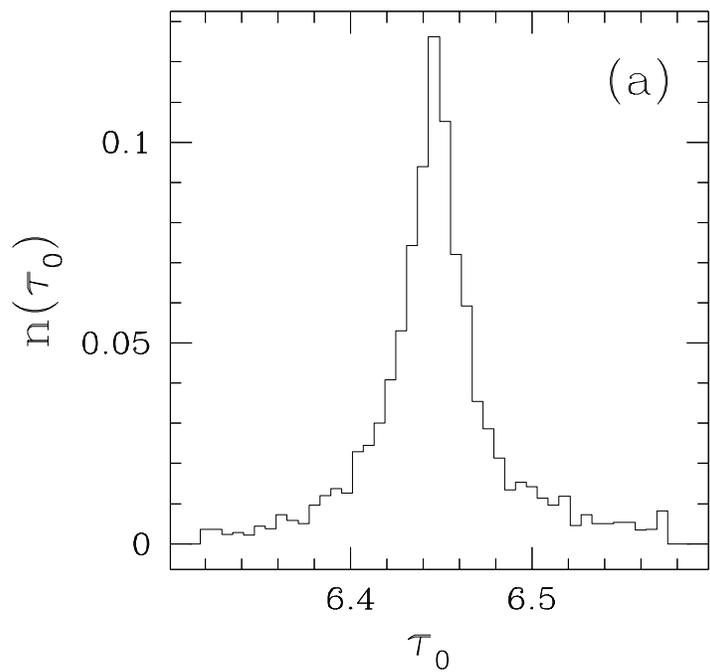
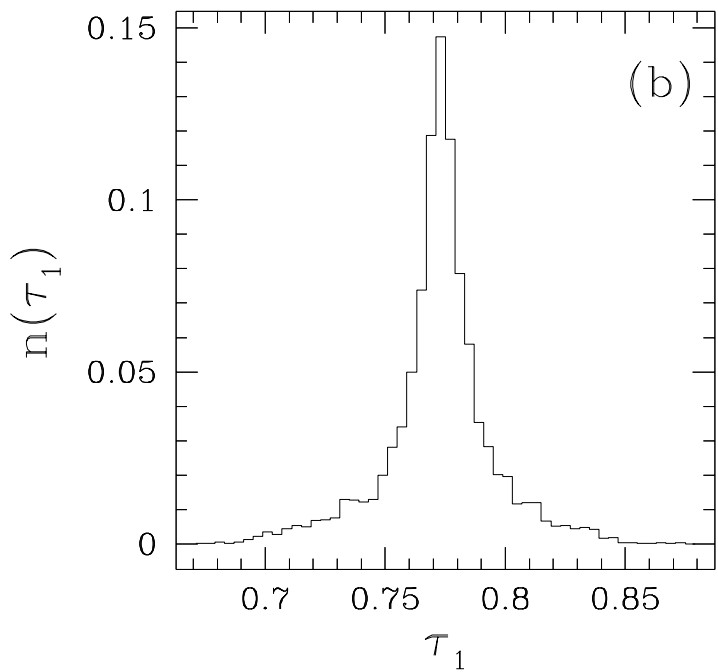
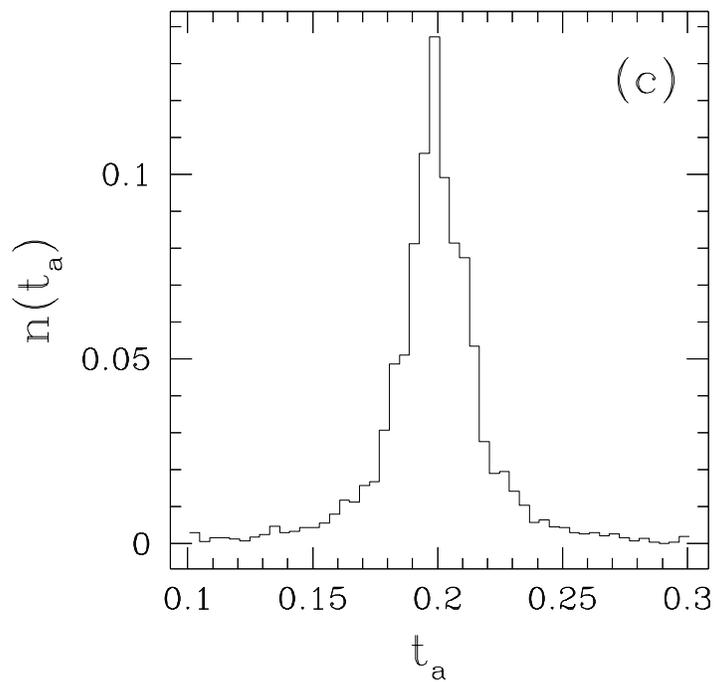
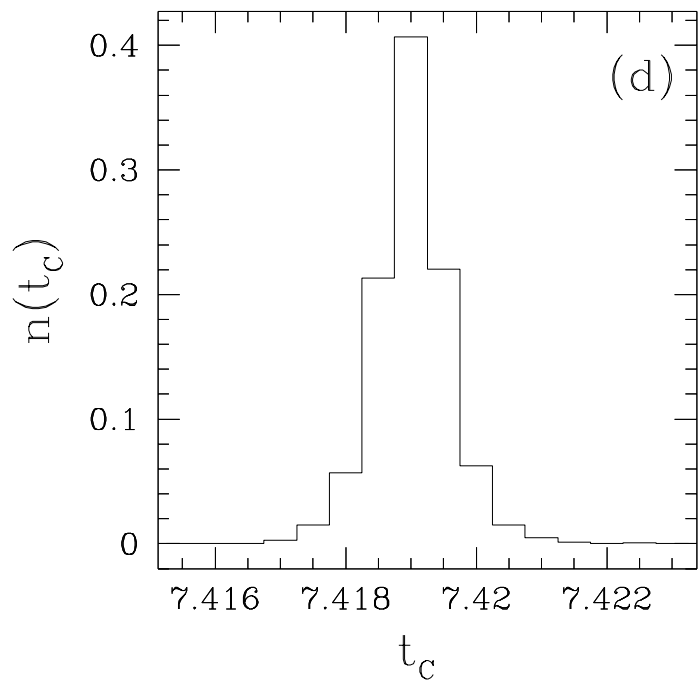
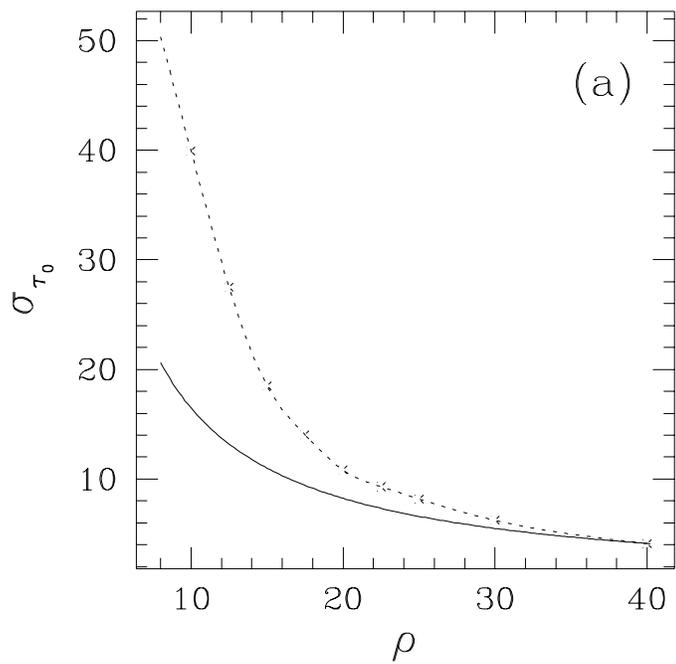
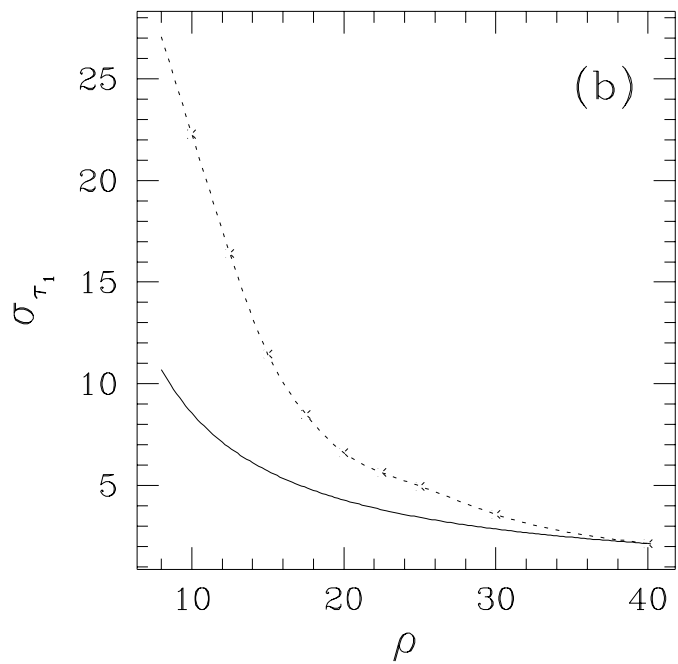
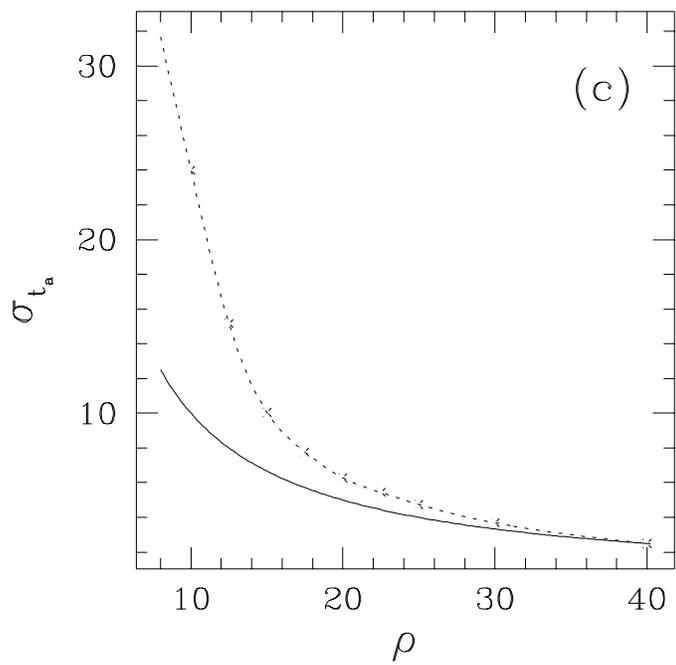
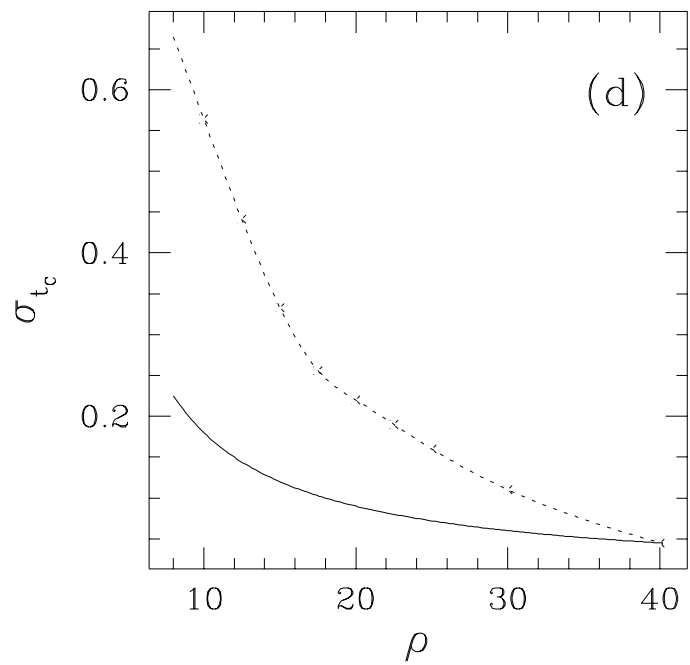
\end{document}